\documentclass{agujournal2019}
\usepackage{url} %this package should fix any errors with URLs in refs.
\usepackage{lineno}
\usepackage{soul}
\usepackage{upgreek}
\draftfalse

\journalname{JGR: Planets}

\begin{document}

%% ------------------------------------------------------------------------ %%
%  Title
%
% (A title should be specific, informative, and brief. Use
% abbreviations only if they are defined in the abstract. Titles that
% start with general keywords then specific terms are optimized in
% searches)
%
%% ------------------------------------------------------------------------ %%

% Example: \title{This is a test title}

%\title{Water ice aerosols in the Martian atmosphere during the 2018
	%global dust storm from TGO/ACS-MIR data}
\title{Martian water ice clouds during the 2018 global dust storm as observed 
	   by the ACS-MIR channel onboard the Trace Gas Orbiter}

%% ------------------------------------------------------------------------ %%
%
%  AUTHORS AND AFFILIATIONS
%
%% ------------------------------------------------------------------------ %%

% Authors are individuals who have significantly contributed to the
% research and preparation of the article. Group authors are allowed, if
% each author in the group is separately identified in an appendix.)

% List authors by first name or initial followed by last name and
% separated by commas. Use \affil{} to number affiliations, and
% \thanks{} for author notes.
% Additional author notes should be indicated with \thanks{} (for
% example, for current addresses).

% Example: \authors{A. B. Author\affil{1}\thanks{Current address, Antartica}, B. C. Author\affil{2,3}, and D. E.
% Author\affil{3,4}\thanks{Also funded by Monsanto.}}

%\authors{Aur\'{e}lien Stcherbinine\affil{1,2}, Mathieu Vincendon\affil{1}, Franck Montmessin\affil{2},
	%Michael J. Wolff\affil{3}, Oleg Korablev\affil{4}, Anna Fedorova\affil{4}}
\authors{A. Stcherbinine\affil{1,2}, M. Vincendon\affil{1}, F. Montmessin\affil{2},
	M. J. Wolff\affil{3}, O. Korablev\affil{4}, A. Fedorova\affil{4}, A. Trokhimovskiy\affil{4},
	A. Patrakeev\affil{4}, G. Lacombe\affil{2}, L. Baggio\affil{2},	A. Shakun\affil{4}}

\affiliation{1}{Institut d'Astrophysique Spatiale, Universit\'e Paris-Saclay, CNRS, 91405 Orsay, France}
\affiliation{2}{LATMOS/IPSL, UVSQ Universit\'e Paris-Saclay, Sorbonne Universit\'e, CNRS, 78280 Guyancourt, France}
\affiliation{3}{Space Science Institute, 4750 Walnut Street, Suite 205, Boulder, Colorado, 80301, USA}
\affiliation{4}{Space Research Institute (IKI), 84/32 Profsoyuznaya, 117997 Moscow, Russia}

%% Corresponding Author:
% Corresponding author mailing address and e-mail address:

% (include name and email addresses of the corresponding author.  More
% than one corresponding author is allowed in this LaTeX file and for
% publication; but only one corresponding author is allowed in our
% editorial system.)

% Example: \correspondingauthor{First and Last Name}{email@address.edu}

\correspondingauthor{Aur\'{e}lien Stcherbinine}{aurelien.stcherbinine@ias.u-psud.fr}

%% Keypoints, final entry on title page.

%  List up to three key points (at least one is required)
%  Key Points summarize the main points and conclusions of the article
%  Each must be 100 characters or less with no special characters or punctuation and must be complete sentences

% Example:
% \begin{keypoints}
% \item	List up to three key points (at least one is required)
% \item	Key Points summarize the main points and conclusions of the article
% \item	Each must be 100 characters or less with no special characters or punctuation and must be complete sentences
% \end{keypoints}

\begin{keypoints}
\item Monitoring of Martian water ice clouds and derivation of vertical profiles of particle size using the 3~$\upmu$m spectral band.
\item Observation of mesospheric water ice clouds at altitudes $\geq$ 90~km during the MY~34 GDS.
\item Evidence of water ice particles $\geq$ 1.5~$\upmu$m between 50~km and 70~km during the GDS.
\end{keypoints}

%% ------------------------------------------------------------------------ %%
%
%  ABSTRACT and PLAIN LANGUAGE SUMMARY
%
% A good Abstract will begin with a short description of the problem
% being addressed, briefly describe the new data or analyses, then
% briefly states the main conclusion(s) and how they are supported and
% uncertainties.

% The Plain Language Summary should be written for a broad audience,
% including journalists and the science-interested public, that will not have 
% a background in your field.
%
% A Plain Language Summary is required in GRL, JGR: Planets, JGR: Biogeosciences,
% JGR: Oceans, G-Cubed, Reviews of Geophysics, and JAMES.
% see http://sharingscience.agu.org/creating-plain-language-summary/)
%
%% ------------------------------------------------------------------------ %%

%% \begin{abstract} starts the second page

\begin{abstract}
The Atmospheric Chemistry Suite (ACS) instrument onboard the ExoMars Trace Gas Orbiter (TGO)
ESA-Roscosmos mission began science operations in March 2018.
ACS Mid InfraRed (MIR) channel notably provides solar occultation observations of the martian atmosphere in the 2.3 -- 4.2~$\upmu$m spectral range.
Here we use these observations to characterize water ice clouds
before and during the MY 34 Global Dust Storm (GDS).
We developed a method to detect water ice clouds with mean particle size $\leq 2~\upmu$m,
and applied it to observations gathered between $L_s=165^\circ$ and $L_s=243^\circ$.
We observe a shift in water ice clouds maximum altitudes from about 60~km before the GDS to above 90~km during the storm.
These very high altitude, small-sized ($r_\mathrm{eff} \leq 0.3~\upmu$m) water ice clouds are more frequent during MY34 
compared to non-GDS years at the same season.
Particle size frequently decreases with altitude, both locally within a given profile and globally in the whole dataset. 
We observe that the maximum altitude at which a given size is observed can increase during the GDS by several tens of km for certain sizes. 
We notably notice some large water ice particles ($r_\mathrm{eff}\geq1.5~\upmu$m) at surprisingly high
altitudes during the GDS (50 -- 70 km).
These results suggest that GDS can significantly impact the formation 
and properties of high altitude water ice clouds as compared to the usual perihelion dust activity.
\end{abstract}

\section*{Plain Language Summary}
In this article, we use data from the Atmospheric Chemistry Suite (ACS) 
infrared spectrometer onboard the European Space Agency (ESA)-Roscosmos Exomars
Trace Gas Orbiter (TGO) mission to study water ice clouds in the
Martian atmosphere. More specifically, we aim to characterize the evolution of their altitude, 
geographic distribution and microphysical properties before and during the planet-wide 
dust storm that occurred during the summer of 2018.
In particular, we developed a method to simultaneously detect the water ice clouds and constrain 
their particle size using simulated spectra of water ice. 
We observe that the maximal altitude of the clouds increased from 60~km to above 90~km during the storm.
Most high altitude clouds have small particle sizes (lower than 0.3~microns) as expected from the low pressure at such altitude.
However, we also observe for the first time large (larger than 1.5~microns) water ice particles at unusually high altitude
(higher than 60~km), uniquely during the storm. This suggests that the increased atmospheric activity associated with global dust 
storm significantly impacts water ice clouds formation.

%% ------------------------------------------------------------------------ %%
%
%  TEXT
%
%% ------------------------------------------------------------------------ %%

\section{Introduction}	\label{sec:intro}

		Since the first spectroscopic detection by Mariner 9 \cite{hanel_1972}, water ice clouds
		have been extensively studied because of their connection to the Martian water cycle 
		\cite{clancy_1996, madeleine_2012, smith_2013, clancy_2017, guzewich_2019}.
		Clouds scatter and absorb incoming solar radiation, thus impacting atmospheric structure and 
		temperature. As a result, water ice cloud particles can modify the global circulation of the Martian atmosphere \cite{wilson_2008}.
		Clouds are also a major actor in the inter-hemispheric water exchange \cite{clancy_1996}. 
		As the evolution of our understanding of the martian climate shows an increasing role of
		water ice clouds,
		there is a growing need to better characterize the properties of water ice
		aerosols in order to better understand and model Mars climate and weather 
		\cite{richardson_2002, montmessin_2004, navarro_2014}.
		Recent studies have notably reveal how more precise vertical representation of water-related atmospheric phenomena
		may impact water cycle modeling \cite{vals_2018}.
		Precise observational constraints about the actual microphysiccal properties of clouds as a function of altitude, 
		notably particle size, are thus of interest to better characterize the whole water cycle.
		
		Planetary-scale storms, characterized by widespread lifting and transport of dust particles, 
		modify the thermodynamics and circulation of the Martian atmosphere on a global scale, 
		and subsequently affect the water cycle. While regional dust storms are recurrent phenomena on Mars, 
		few of them evolve into events that encircle the whole planet
		to become global dust storms (hereafter "GDS"). 
		Such events are erratic, with an average of one occurrence every three to four Martian years
		\cite{zurek_1993, clancy_2000, kass_2016}. 
		The first two GDSs of the 21$^\mathrm{th}$ century occurred 6 Earth years 
		apart, in 2001 and 2007, corresponding to Martian Years (hereafter "MY") 25 and 28
		\cite<e.g.>{wang_richardson_2015}.
		Two large (regional) dust storms developed in 2018 (MY 34): one in the Northern
		hemisphere starting at $L_s~=~181^\circ$, and the other in the Southern hemisphere at
		$L_s~=~188^\circ$. 
		They subsequently merged to become a GDS at $L_s~=~193^\circ$, lasting 
		until $L_s~=~250^\circ$ \cite{guzewich_GDS_2019, sanchez-lavega_2019, smith_2019}.
		This GDS occured near the Northern fall equinox, however, it should be noted that there is no unique timing for the onset and development of GDSs on Mars.
		For instance, the MY 28 GDS, occured significantly later, at the end of the so-called storm season (from $L_s=260^\circ$ to $L_s=310^\circ$)
		\cite{wang_richardson_2015}. This variability implies that a variety of scenarios dictate the formation of GDSs, 
		requiring individual assessments of GDS's.
		
	    The MY 28 dust storm has been shown to have profoundly influenced the water vapor atmospheric
	    distribution, elevating water well above the troposphere ($>$60~km) \cite{fedorova_2018}
	    while pushing up the hygropause altitude \cite{heavens_2018}.
		A rapid increase of the H$_2$O and HDO abundances at altitudes between 40 and 80~km
		\cite{nature_hdo} was observed during MY 34 storm, and recently reproduced in a GCM
		simulation \cite{neary_2019}.
		This large augmentation of the high altitude water content is believed to boost
		hydrogen, or nearly equivalently, water escape from Mars \cite{chaffin_2017, heavens_2018, fedorova_2020}.
		However, the formation of clouds impacts the ability of water (or hydrogen) to escape as it confines (in theory) a significant fraction of water below the cloud level.
		Thus, understanding the extent to which high altitude water vapor condenses to form clouds 
		during GDSs is important in assessing the possible propagation pathways 
		of water vapor up to the exobase \cite{neary_2019}. 
		Because water vapor can exist in a supersaturated state in the Martian atmosphere \cite{maltagliati_2011}
		and has been observed in a large amount during the MY 34 GDS \cite{fedorova_2020},
		simply supposing that the freezing point can be used to set the limits where water is free to be mobilized is likely to be problematic. 
		The availability of condensation nuclei, supplied by the ubiquitous atmospheric dust particles, and possibly by interplanetary dust particles and micrometeoric smoke that aggregates as entering the Martian atmosphere \cite{crismani_2017, plane_2018, hartwick_2019}, is another key factor controlling the potential for water ice clouds. It has for example been shown by a number of studies \cite{michelangeli_1993} that the occurrence of supersaturation is anti-correlated to the availability of condensation nuclei. 
		The	condensation nuclei themselves are expected to be much more abundant during GDS, including with larger sizes
		\cite<e.g.>{wolff_2003, clancy_2010, lemmon_2019}, which may impact how and when water ice clouds form. 
		It is however difficult to predict the distribution and properties of condensation nuclei at high altitude during GDS, which additionally modify the thermal structure of the atmosphere. In this complex situation, accumulating observational constraints about the actual behavior of water ice at high altitude during GDS is required to provide further constraints on the fate of high altitude water
		vapor during these events. 
		
		The ExoMars Trace Gas Orbiter (TGO), ESA-Roscosmos Atmospheric Chemistry Suite (ACS) 
		and Nadir and Occultation for MArs Discovery (NOMAD) instruments 
		started their science operation phase in March 2018 
		\cite{korablev_ACS, vandaele_nomad_2018, nature_methane, nature_hdo}, before the MY 34 GDS.
		The ACS middle-infrared (MIR) channel is a crossed-dispersion echelle spectrometer dedicated to 
		solar occultation (hereafter, "SO"): each observation covers a $\sim$300~nm wide spectral interval
		selected between 2.3 and 4.2~$\upmu$m. 
		This interval is set by rotating the spectrometer secondary grating (i.e., the diffraction order separation) 
		to align the interval of interest with the detector. The cross-dispersion aspect provides from 10 to 21 
		diffraction orders (spectral segments) stacked on top of each other. 
		The separation between consecutive orders is only dependent on the number of orders displayed at once and 
		thus on the secondary grating position. 
		Thirteen different positions can be employed to completely sample the full accessible spectral range of the instrument.
		With the two hours orbital period of TGO and the performing of SO in the mid-infrared, ACS and NOMAD provides a totally new and huge dataset of vertical profiles of the atmospheric extinction \cite{korablev_ACS, vandaele_nomad_2018}.
		We are using the ACS-MIR channel to study the extinction properties of the Martian airborne 
		particles in the 3~$\upmu$m spectral region, which possesses a distinct	diagnostic 
		capability to identify the O-H stretching signature; whether it is due to water ice absorption 
		or bound water in dust. In the case of water ice, the depth and shape of the feature depend on both
		the abundance and the particle size of water ice \cite{vincendon_2011, guzewich_2014, clancy_2019}.
		The ACS-MIR occultation data in this spectral range gives us complementary information to 
		those derived from previous limb and nadir scattered light observations by OMEGA
		\cite{bibring_OMEGA_2004, vincendon_2011} and CRISM 
		\cite{murchie_CRISM_2007, clancy_2019, guzewich_2014}.
		We focus here on the use of the water ice absorption feature during the MY 34 GDS that offer a relevant framework  
		for a further characterization of the interactions between water and dust
		during extreme dust events.

\section{Data analysis}	\label{sec:data_analysis}
\begin{figure}[h!]
			\centering
			\includegraphics[width=\textwidth]{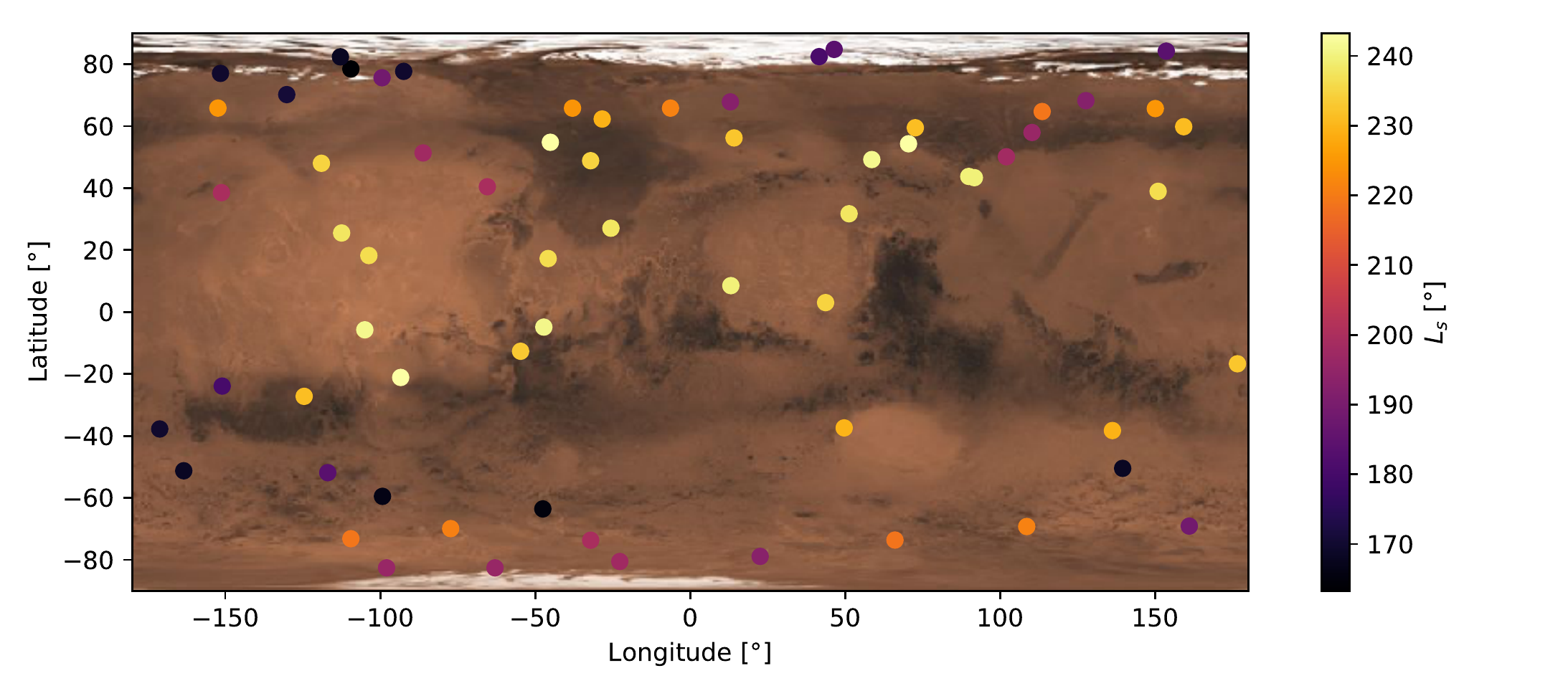}
			\caption{Spatial and time distribution of the 65 ACS-MIR  observations used in this study.}
			\label{fig:distrib_lat_lon_ls}
		\end{figure}

		In order to study the 3~$\upmu$m water ice absorption band, we use the instrument configuration centered on the 3.1 -- 3.4~$\upmu$m spectral range corresponding to the 12th secondary grating position.
		Figure~\ref{fig:distrib_lat_lon_ls} presents the spatial and temporal
		distribution of the position~12 MIR observations used in this study, covering a temporal
		range from $L_s~=~165^\circ$ to $L_s~=~243^\circ$. 
		However, due to the SO geometry, observations only occur in the periods near local times of 06:00AM and 06:00PM.

	\subsection{Continuum extraction}	\label{sec:continuum}
\begin{figure}[h!]
			\centering
			\includegraphics[width=\textwidth]{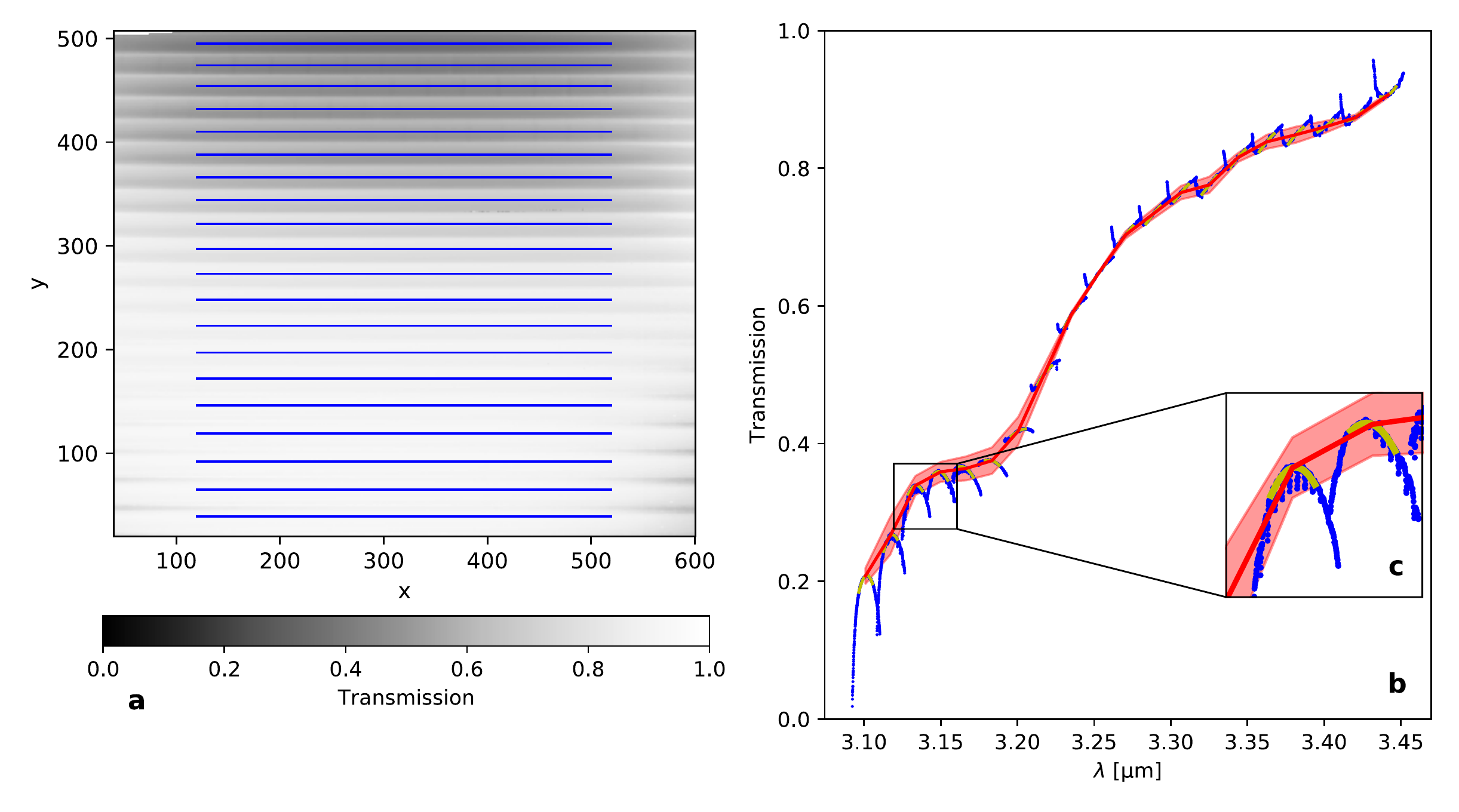}
			\caption{Extraction of the spectral continuum. 
			    \textbf{a.} Reduced calibrated image of the ACS-MIR detector containing
				20 diffraction orders along the y-axis. For each order, the wavelength varies
				along the x-axis. Each \textcolor{blue}{blue} line corresponds to the detector
				pixels used to obtain the spectrum of one diffraction order.
				\textbf{b.} Corrections steps for the extraction of the spectral continuum.
				In \textcolor{blue}{blue} the initial data, in \textcolor{yellow!90!black}{yellow}
				the polynomial fit of the 200 points from the center of each order (after the
				third iteration),
				and in \textcolor{red}{red} the extracted continuum (with the associated 
				uncertainties in the red shadowed region).
				\textbf{c.} Zoom on two diffraction orders.}
			\label{fig:etapes_corr}
		\end{figure}

		Every ACS-MIR observation is composed of $\sim$20 spectral segments individually exhibiting broad instrumental shapes that generally prevents contiguous segments from matching precisely in their overlapping wavelength intervals (cf. figure~\ref{fig:etapes_corr}) due to instrumental effects \cite{korablev_ACS,trokhimovskiy_2015}. As the aerosol property derivation is intrinsically dependent of the continuum behavior, we have adopted a method to extract the continuum across all displayed fraction orders, taking into account this spectral shape of each order. This allows us to derive the continuum shape across the 3 to 3.4 $\upmu$m range.

        In order to do this, we have chosen to fit the 200 centered spectral points of each diffraction order with
        a 2$^\mathrm{nd}$ degree polynomial, and use an iterative method to remove the gas absorption bands
        that tend to bias the continuum values represented by the fitted polynomial.
        After the first iteration, for each diffraction order we determine the standard deviation of the 
        transmission values of the considered spectral points $\sigma$, and we consider that all the points
        with transmission lower than the polynomial fit by at least $\sigma$
        (i.e. $\mathrm{Tr}(\lambda) < (\mathrm{fit}(\lambda) - \sigma)$) are affected by some gas
        absorption. So we remove these points as they tend to underestimate the actual continuum.
        In a second step, we perform the same filtering method from a polynomial fit 
        on each diffraction order for the remaining wavelengths 
        (i.e. the 200 centered points without the ones with transmission values lower than 
        ($\textrm{fit} - \sigma$)).
        Finally, we perform a third fit to the continuum and then only retain the fitted transmission value 
        at the center point of each fitted order (cf. figure~\ref{fig:etapes_corr}.b).
        As a result, the complete spectrum has an effective spectral resolution of one point per diffraction order,
        which corresponds to a spectral resolution of $\sim 1-2$~nm.  This is a much smaller number of points than the ACS-MIR native sampling, but is sufficient to capture the general shape of the continuum used in the next sections for solid particles characterisation.
        Next, we estimate the uncertainties of the data as the sum of the standard deviation of the difference between the data and the polynomial fit for each order (\emph{random uncertainties}), and the maximal gap between the polynomial fit of the order and the extracted continuum (1 point per order), linearly interpolated between each diffraction order
        (\emph{systematic uncertainties} due to the curvature of the diffraction orders, see figure~\ref{fig:etapes_corr}.c).
		Additionally, in the following we do not consider the outer regions of the spectra in order to
		avoid detector edge effects, i.e. the spectels corresponding to 3.10, 3.12 and 3.44~$\upmu$m.
		
    \subsection{Haze top determination}
\begin{figure}[h!]
		    \centering
		    \includegraphics[width=\textwidth]{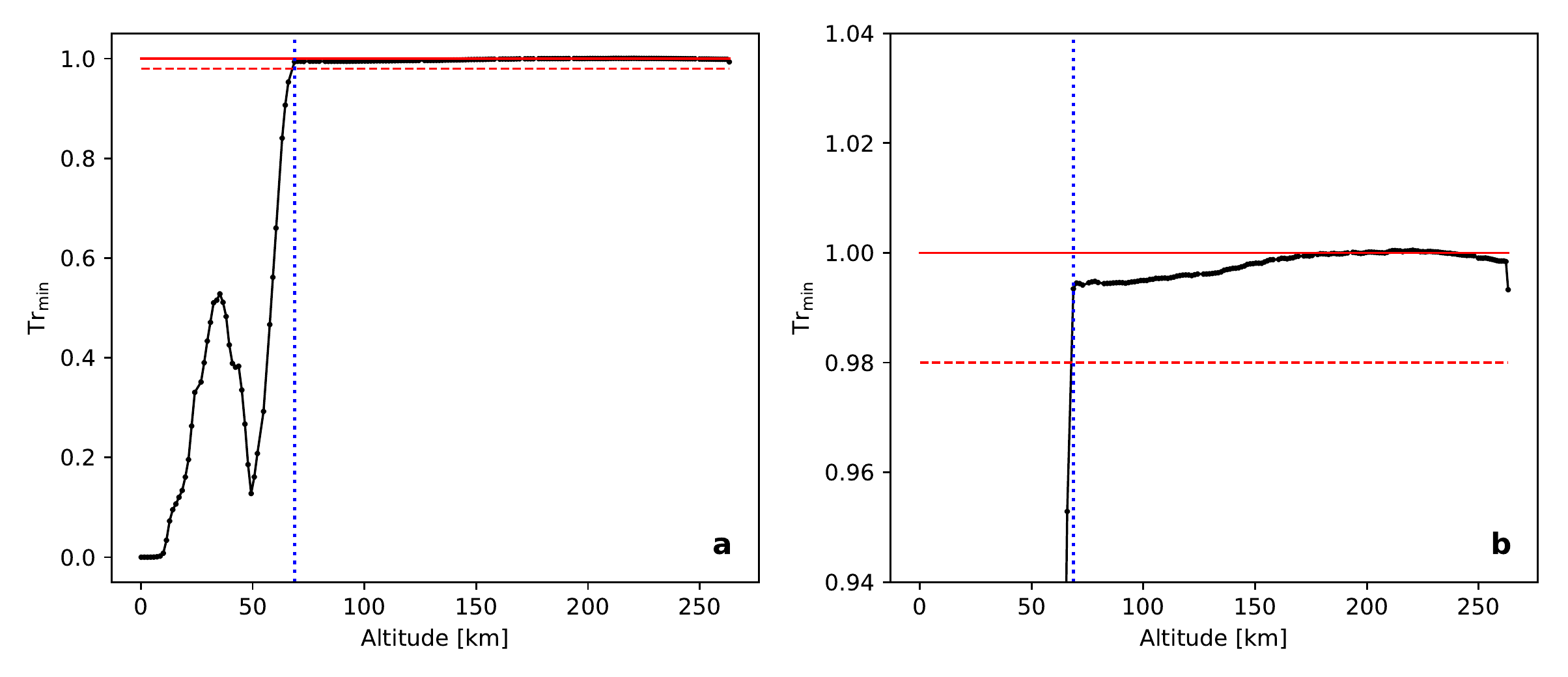}
		    \caption{Haze top determination. \textbf{a.} Example of transmission spectra
		        as a function of altitude. The \textcolor{red}{red} lines correspond respectively
		        to transmission levels of 1 (solid line) and 0.98 (dashed line).
		        \textbf{b.} Zoom on the plateau reached for transmissions greater than about 0.99. A conservative threshold of 0.98 corresponding to $\varepsilon=0.02$ has been selected to calculate the haze top altitude shown by the \textcolor{blue}{blue} dotted line (69 km in this example).}
		    \label{fig:seuil_htop}
		\end{figure}

		Using this extracted continuum at each observed tangent point, we can determine the haze top
		altitude for each observation. We define haze top as the highest altitude at which aerosols can be sensed along the line of sight. We calculate haze top
		by finding the first altitude for which the transmission is greater
		than a defined threshold ($1 - \varepsilon$).
		$\varepsilon$ introduces the consideration of the artificial non-steady nature of the measured transmission as it approaches unity at high altitude. We observe empirically that this transition occurs above transmissions around 0.99. We have thus selected a conservative value of $\varepsilon=0.02$ for this threshold. An example is provided in the figure~\ref{fig:seuil_htop}.
		
    \subsection{3 $\upmu$m band monitoring}	\label{sec:clouds}
\begin{figure}[h!]
			\centering
			\includegraphics[width=10cm]{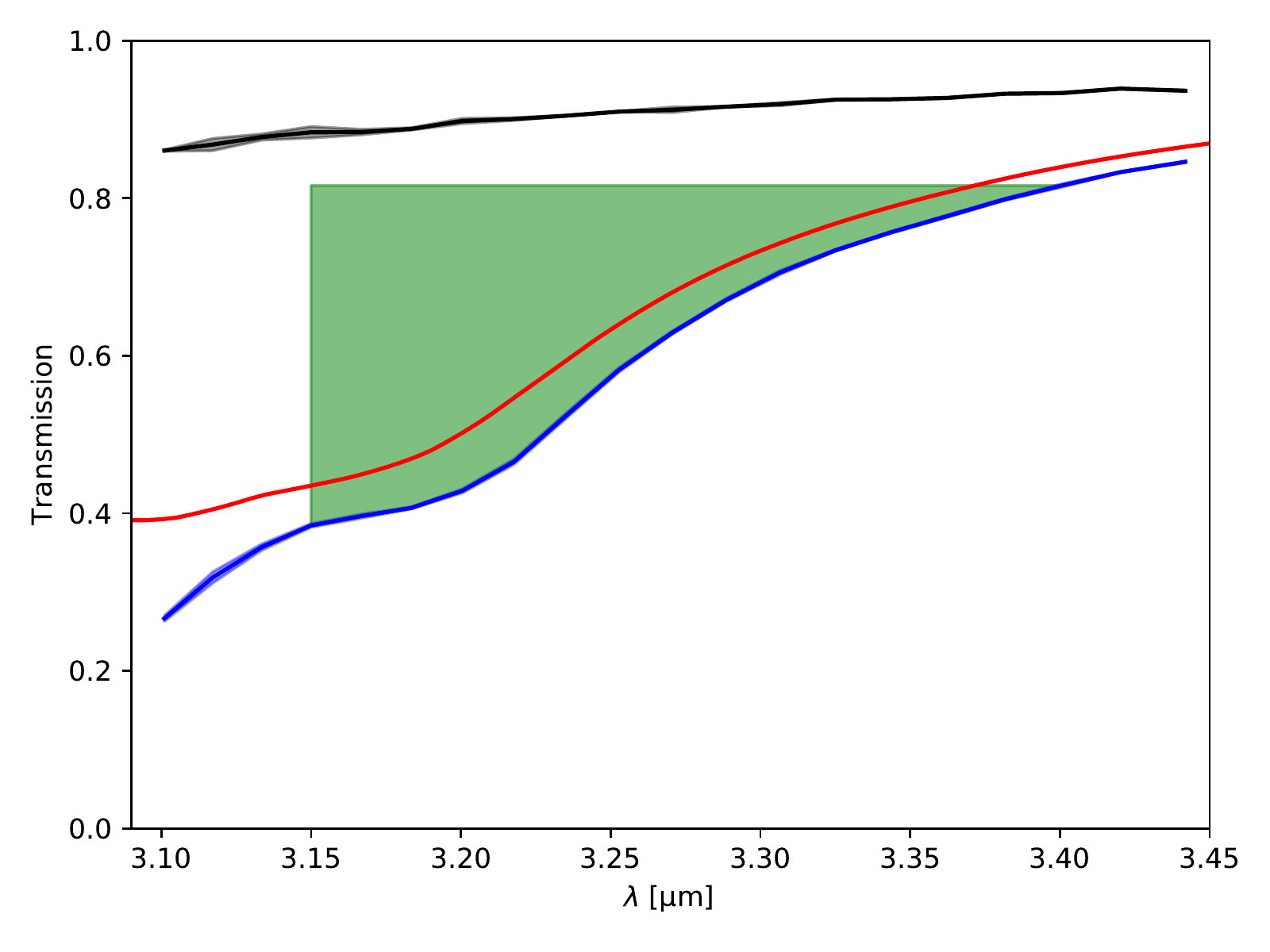}
			\caption{Calculation of the 3~$\upmu$m Integrated Band Depth (IBD) which can be used as a first order proxy for the presence of some water ice clouds. In \textcolor{blue}{blue}, we show an example of ACS-MIR position 12 spectra, showing a strong 3~$\upmu$m absorption band.
				In \textcolor{green!60!black}{green} is shown the measurement of the 3~$\upmu$m
				absorption using the IBD method (cf. equation (\ref{eq:IBD})) with
				$\lambda_1 = 3.15~\upmu$m and $\lambda_2 = 3.4~\upmu$m). This spectrum is compared to the theoretical transmission
				of 0.5~$\upmu$m spherical water ice  particles in \textcolor{red}{red} (see section~\ref{sec:model_Mike}) and in black, another example of ACS-MIR spectra with a low 3~$\upmu$m band, and presumably no water ice.}
			\label{fig:theory_3microns}
		\end{figure}

		The 3~$\upmu$m absorption feature corresponds to the OH/H$_2$O absorption. We quantify the depth of this feature as a function of 
		altitude
		using an Integrated Band Depth (IBD) method \cite{jouglet_2007, calvin_1997}. 
		This approach computes the mean depth of transmission between two selected wavelengths 
		(see eq.~\ref{eq:IBD} and figure~\ref{fig:theory_3microns}). 
		
		\begin{linenomath*}
		\begin{equation}	\label{eq:IBD}
			\mathrm{IBD}(\mathrm{Tr}, \lambda_1, \lambda_2) = 
				\frac{1}{\lambda_2 - \lambda_1} \int_{\lambda_1}^{\lambda_2}
				\left[ \mathrm{Tr}(\lambda_2) - \mathrm{Tr}(\lambda) \right] d\lambda
		\end{equation}
		\end{linenomath*}

	\subsection{Extinction coefficient determination}	\label{sec:inv_vert}
		In order to retrieve the extinction profile and properties of the aerosols, we employ a
		vertical inversion algorithm based on the onion-peeling method \cite{goldman_saunders_1979},
		from the total optical depth $\tau$ (i.e., as integrated along the line of sight). The result is the the extinction coefficient $k_\mathrm{ext}$ as a function of altitude.
		Then, we can compute the opacity of a cloud by integrating the extinction coefficient along a vertical
		line between the boundary altitudes detected for the cloud (see the following sections for a description of the water ice
		clouds identification).
		
		For each altitude level $i$, the optical depth is given by:
		\begin{linenomath*}
		\begin{equation}	\label{eq:tau_scalaire}
			\tau_i(\lambda) = \sum_{j=i}^{N} 
				\underbrace{2\times\left(\sqrt{R_{j+1}^2 - R_i^2} - \sqrt{R_j^2 - R_i^2}\ \right)}%
															_{A_{i, j}} k_{\mathrm{ext}, j}(\lambda)
		\end{equation}
		\end{linenomath*}

		That we can rewrite in a matrix manner:
		\begin{linenomath*}
		\begin{equation}	\label{eq:tau_matrice}
			\tau = A\, k_\mathrm{ext}
		\end{equation}
		\end{linenomath*}

		The transfer matrix $A$ is upper triangular so it is 
		invertable and we can write :
		\begin{linenomath*}
		\begin{equation}	\label{eq:kext_matrice}
			k_\mathrm{ext} = A^{-1}\, \tau
		\end{equation}
		\end{linenomath*}

		I.e.
		\begin{linenomath*}
		\begin{equation}	\label{eq:kext_scalaire}
			k_{\mathrm{ext}, i} = \sum_{j=i}^{N} \left({A^{-1}}\right)_{i, j}\ \tau_j
		\end{equation}
		\end{linenomath*}

		Then, according to \citeA{JCGM_100}, as $\tau = -\log(\mathrm{Tr})$ and $k_\mathrm{ext}$ 
		given as a function of $\tau$ in (\ref{eq:kext_scalaire}), we have the uncertainties on the
		optical depth values (\ref{eq:error_tau}) and their propagation to the extinction 
		coefficient $k_\mathrm{ext}$ throughout the vertical inversion given by 
		(\ref{eq:error_kext}).

		\begin{linenomath*}
		\begin{equation}	\label{eq:error_tau}
			\Delta \tau(\lambda) = \frac{\Delta \mathrm{Tr}(\lambda)}{\mathrm{Tr}(\lambda)}
		\end{equation}
		\end{linenomath*}

		\begin{linenomath*}
		\begin{equation}	\label{eq:error_kext}
			\Delta \left(k_{\mathrm{ext}, i}\right) = 
				\sqrt{\sum_{j=i}^{N} \left(A^{-1}\right)_{i, j}^2	\left(\Delta \tau_j\right)^2}
		\end{equation}
		\end{linenomath*}
		
		\begin{notation}
			\notation{$N$} The number of atmospheric layers.
			\notation{$\mathrm{Tr}_i$} The measured transmission value at the $i^\mathrm{th}$
				altitude.
			\notation{$\tau_i$} The optical depth observed at the $i^\mathrm{th}$ altitude
				(integrated along the line of sight).
			\notation{$k_{\mathrm{ext}, i}$} The extinction coefficient of the $i^\mathrm{th}$ 
				atmospheric layer.
			\notation{$R_i$} The radius of the sphere corresponding to the bottom altitude of the 
				$i^\mathrm{th}$ atmospheric layer, i.e. the altitude of the bottom of the layer 
				added to the Martian radius.
		\end{notation}

		\begin{figure}[h!]
			\centering
			\includegraphics[width=15cm]{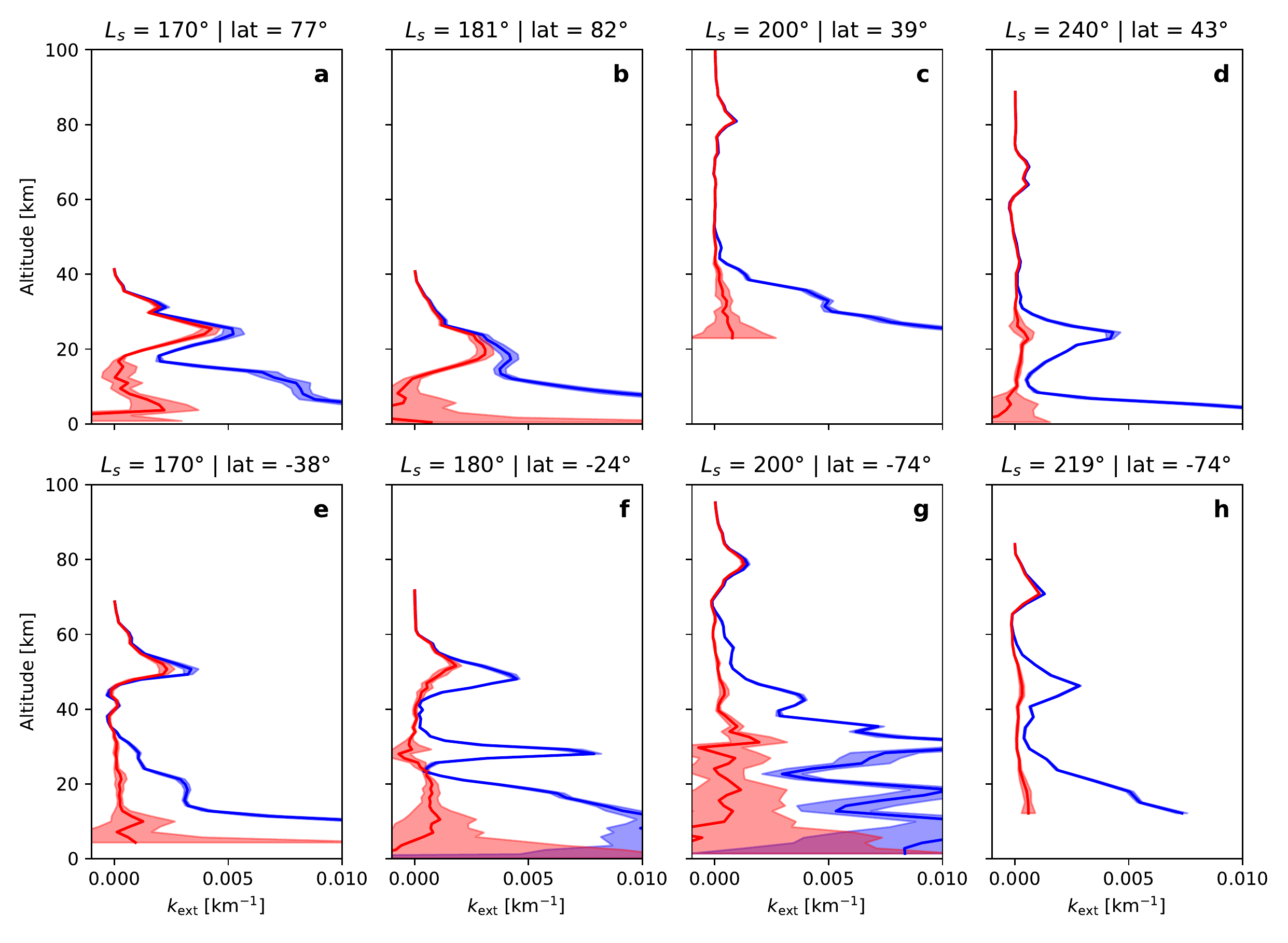}
			\caption{Typical extinction coefficient vertical profiles retrieved from different ACS-MIR 
				observations.
				In \textcolor{blue}{blue} the extinction coefficient at 3.2~$\upmu$m, which provides a proxy for the overall particle (dust and ice) profile. In
				\textcolor{red}{red} the difference between the extinction coefficient at 3.2~$\upmu$m
				and 3.4~$\upmu$m, which provides a proxy for the presence of band depth associated with small-grained water ice We can see layers with various behavior, some typical of the presence of small-grained water ice (e.g., e at about 50~km) and some interpreted as resulting from either large-grained water ice, or dust (e.g., d at 20~km). See section~\ref{sec:results} for discussion. 
				The shadows regions show the uncertainties on the $k_\mathrm{ext}$ values.}
			\label{fig:kext_vert}
		\end{figure}

		Figure~\ref{fig:kext_vert} presents several examples of retrieved vertical profiles of the extinction
		coefficient $k_\mathrm{ext}$ from ACS-MIR observations before and
		during the MY 34 GDS. We can already notice the presence of one or several detached layers in some profiles, which are located at various altitudes. We also observe strong spectral variability between observations, with some layers having a distinctive signature when comparing the transmissions at 3.2 and 3.4~$\upmu$m (compare e.g. blue and red profiles in panels a and d of figure~\ref{fig:kext_vert}). We will discuss in the next section how this spectral behavior can be modeled to derive information about the presence and properties of water ice clouds.

\section{Water ice cloud modeling}	\label{sec:clouds_prop}
	\subsection{Model and hypothesis}	\label{sec:model_Mike}
\begin{figure}[h]
			\centering
			\includegraphics[width=\textwidth]{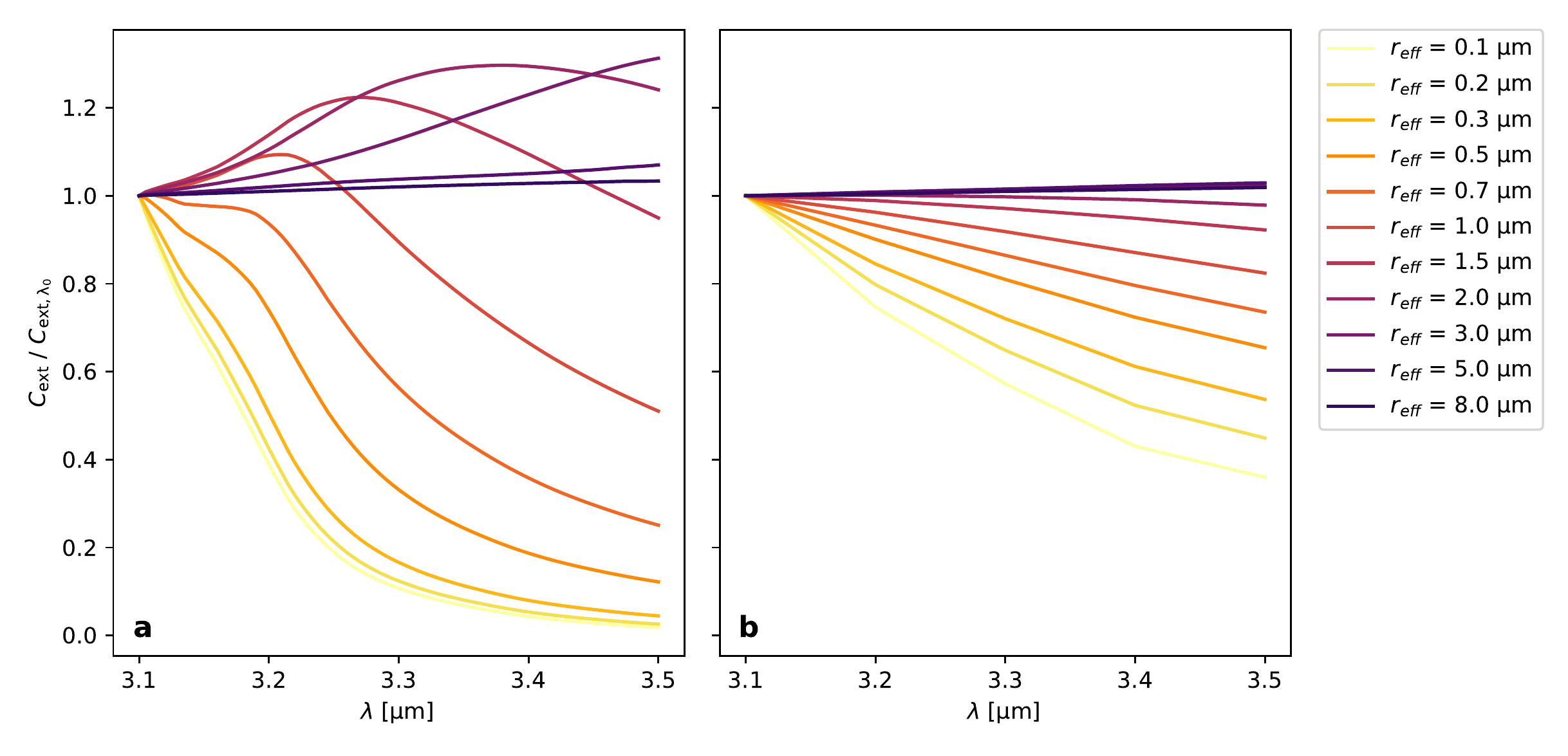}
			\caption{Water ice and dust models. \textbf{a.}
				Simulated extinction opacity spectra $C_\mathrm{ext}(\lambda)$ for water ice 
				spherical particles of different sizes, normalized at 
				$\lambda_0~=~3.1~\upmu\textrm{m}$.
				\textbf{b.} Same plot for dust spherical particles.
				For particles with $r_\mathrm{eff} > 3~\upmu$m, we cannot distinguish between water
				ice and dust signature in this spectral window.}
			\label{fig:models_Mike}
		\end{figure}

		In order to identify water ice clouds, and constrain their particle sizes, we have computed the
		wavelength dependence of the extinction coefficient of pure water ice and dust layers of various particle sizes (figure~\ref{fig:models_Mike}). These extinction coefficients are calculated using a public domain
		Mie code \cite{toon_1981} and assuming a gamma size distribution \cite{hansen_1974}
		with an effective variance of 0.1 \cite<e.g.>[and references contained within]{wolff_2017}.
		
		We can see in the figure~\ref{fig:models_Mike} that the spectral properties of water ice and dust differ significantly in most cases, and that there is a strong dependence of the spectral shape on particle size for water ice. This is not true when water ice particles sizes become greater than about $3~\upmu$m: for such sizes the absorption become quite flat in the 3.1 -- 3.5~$\upmu$m
		range whatever the particle size and it is even not possible to differentiate water ice from dust.
		Particles size lower than 0.1~$\upmu$m have similar $k_\mathrm{ext}$ spectra corresponding to the Rayleigh diffusion regime. As a consequence, we do not make any distinction
		between water ice particles with radii of 0.1~$\upmu$m and lower.
		
		In the following, we fit our retrieved extinction coefficient from ACS data to these models for pure water ice or dust layers in order to identify and characterize water ice rich layers composed of particles lower than $\sim 2~\upmu$m, as results suggesting larger water ice particles are eliminated in the filtering process due to possible confusion with dust (see section~\ref{sec:filtering}). Layers not identified as such can correspond to either dust-dominated or large-grained water ice layers.

	\subsection{Water ice particle size retrieving method}	\label{sec:part_size}
		As the extinction opacity $C_\mathrm{ext}$ is proportional to $k_\mathrm{ext}$,
		we try to fit our ACS-MIR $k_\mathrm{ext}$ data with the modeled
		$C_\mathrm{ext}$ spectra. 
		To do this, we consider the extinction coefficient as a function of the wavelength 
		$\lambda$, the effective radius $r_\mathrm{eff}$ and a scalar factor $\alpha > 0$
		(cf. equation~(\ref{eq:Cext_kext_2})), where $\alpha$ is related to the number 
		of particles that have scattered the light and $C_\mathrm{ext}^\mathrm{interp}$ is
		interpolated from the input models, using a bivariate spline approximation.
		The distinction between $k_\mathrm{ext}$ and $C_\mathrm{ext}$ is just a matter of 
		normalization 
		$\left(\frac{C_\mathrm{ext}(\lambda, r_\mathrm{eff})}{C_\mathrm{ext}(\lambda_0, r_\mathrm{eff})}=
		\frac{k_\mathrm{ext}(\lambda, r_\mathrm{eff})}{k_\mathrm{ext}(\lambda_0, r_\mathrm{eff})}\right)$.
		However, we decide to leave $\alpha$ as a free parameter to get rid of the choice of 
		normalization wavelength $\lambda_0$.

		\begin{linenomath*}
		\begin{equation}	\label{eq:Cext_kext_2}
			k_\mathrm{ext}(\lambda, r_\mathrm{eff}, \alpha) = 
								\alpha\, C_\mathrm{ext}^\mathrm{interp}(\lambda, r_\mathrm{eff})
		\end{equation}
		\end{linenomath*}
		
        Then, in order to avoid the problem of the presence of local minima for $r_\mathrm{eff}$ in the
        \emph{least-square} fitting algorithm, we generate models on a grid of radii of
        0.01~$\upmu$m between 0.1~$\upmu$m and 8~$\upmu$m, and for each one we retrieve the optimal value of $\alpha$
        using the \emph{Trust Region Reflective algorithm} implemented in the 
        \texttt{scipy.optimize.curve\_fit()} Python function \cite{scipy_preprint}.
        Thus, we obtain the best fit for each particle size, that we compare in a second time,
        by computing a $\chi^2$ from the data and the model, as defined in the 
        equation~(\ref{eq:chi2}) \cite{bevington_1992}.
        This allows us to find the global minimum of $\chi^2$ as a function of the particle
        size, and derive the associated $r_\mathrm{eff}$ with this "best fit".
        
        \begin{linenomath*}
        \begin{equation}    \label{eq:chi2}
            \chi^2(r_\mathrm{eff}) =
                    \sum_{i=1}^N \frac{\left(\textrm{data}_i - \textrm{model}_{r_\mathrm{eff},\,i}\right)^2}
                                            {\sigma_i^2}
        \end{equation}
        \end{linenomath*}
        
        Finally, we use the reduced chi-square $\chi^2_\nu$ defined in the equation~(\ref{eq:chi2_red}),
        as a measure of the goodness of the fit \cite{bevington_1992}.
        In theory, a model is considered to be a good approximation of the data when $\chi^2_\nu\le1$.
        
        \begin{linenomath*}
        \begin{equation}    \label{eq:chi2_red}
            \chi^2_\nu(r_\mathrm{eff}) = \frac{1}{N - 2}
                    \sum_{i=1}^N \frac{\left(\textrm{data}_i - \textrm{model}_{r_\mathrm{eff},\,i}\right)^2}
                                            {\sigma_i^2}
                    = \frac{\chi^2}{\nu}
        \end{equation}
        \end{linenomath*}
        
        \begin{notation}
            \notation{$\mathrm{data}_i$} The $i^\mathrm{th}$ spectel of the $k_\mathrm{ext}$ spectra
                from the ACS-MIR observation.
            \notation{$\mathrm{model}_{r_\mathrm{eff},\,i}$} The $i^\mathrm{th}$ spectel of the model 
                extinction spectra for a particle size of $r_\mathrm{eff}$.
            \notation{$\sigma_i$} The uncertainty on the value of data$_i$.
            \notation{$N$} The number of spectral points in the considered spectrum.
            \notation{$\nu$} The number of degrees of freedom ($\nu = N - p$).
            \notation{$p$} The number of fitting parameters (here $p = 2$).
        \end{notation}
        
        Then, in order to quantify the uncertainties for the retrieved particles size, we search for the 
        $r_\mathrm{eff}$ of all the models that will pass our filtering criteria 
        (cf section~\ref{sec:filtering}) during the fitting process of an ACS-MIR spectrum 
        and verify 
        $\chi^2_{\nu,\, \mathrm{ice}} \leq \left(2 \chi^2_{\nu,\, \mathrm{ice}, \, \mathrm{min}} + 1\right)$,
        where $\chi^2_{\nu,\, \mathrm{ice}, \, \mathrm{min}}$ is the minimal value of $\chi^2_{\nu,\, \mathrm{ice}}$ (i.e. associated with the optimal $r_\mathrm{eff}$ value). This last criteria 
        that essentially affects low $\chi^2_{\nu,\, \mathrm{ice}}$ values was
        add to take into account the goodness of the optimal fit in the uncertainties estimation
        As all these models can be considered as acceptable, they will define the uncertainties range of the fit.
        In other words, the optimal $r_\mathrm{eff}$ corresponds to the model with the lower
        $\chi^2_{\nu,\, \mathrm{ice}}$ value, and the lower (respectively upper) bounds for the particle
        size uncertainties corresponds to the minimal (respectively maximal) value of $r_\mathrm{eff}$ in
        the set of models that prove equation~(\ref{eq:critere_filtrage}) along with
        $\chi^2_{\nu,\, \mathrm{ice}} \leq \left(2 \chi^2_{\nu,\, \mathrm{ice}, \, \mathrm{min}} + 1\right)$.
    
	\subsection{Filtering of dust/ice ambiguous cases}  \label{sec:filtering}
		We identify water ice clouds by correlating extinction coefficient spectral behavior compatible with water ice, and incompatible with dust. 
		To do that, we experimentally determined $\chi^2$ thresholds that we present in the following.
		Thus, to be classified unambiguously as a water ice cloud detection, a layer needs to have a $\chi^2_\nu \leq 9$ for water ice, but also a lower quality fit to dust. 
		We then apply the previously described fitting algorithm also for dust and not only for water ice.
		We eliminate all those observations that the water ice models don't improve the $\chi^2_\nu$ by at least a factor~4, as well as the ones that it is possible to fit a dust spectra with a $\chi^2_\nu \leq 1$. 
		That is to say, we only consider the fits which verify the condition (\ref{eq:critere_filtrage}).
		This last criteria was added as a safeguard and is relevant in the case of a fit associated to a low $\chi^2_\nu$ value, where
		even if $\chi^2_{\nu\, \mathrm{dust}} \geq 4\chi^2_{\nu\, \mathrm{ice}}$, the dust model still provide
		a good fit of the data that we cannot ignore.
		
		\begin{linenomath*}
		\begin{equation}    \label{eq:critere_filtrage}
    		\left(\chi^2_{\nu,\, \mathrm{ice}} \leq 9\right) \&
    		\left(\chi^2_{\nu,\, \mathrm{ice}} \leq \frac{\chi^2_{\nu,\, \mathrm{dust}}}{4}\right) \&
    		\left(\chi^2_{\nu,\, \mathrm{dust}} > 1\right)
		\end{equation}
		\end{linenomath*}
		
\begin{figure}[h!]
		    \centering
		    \includegraphics[width=\textwidth]{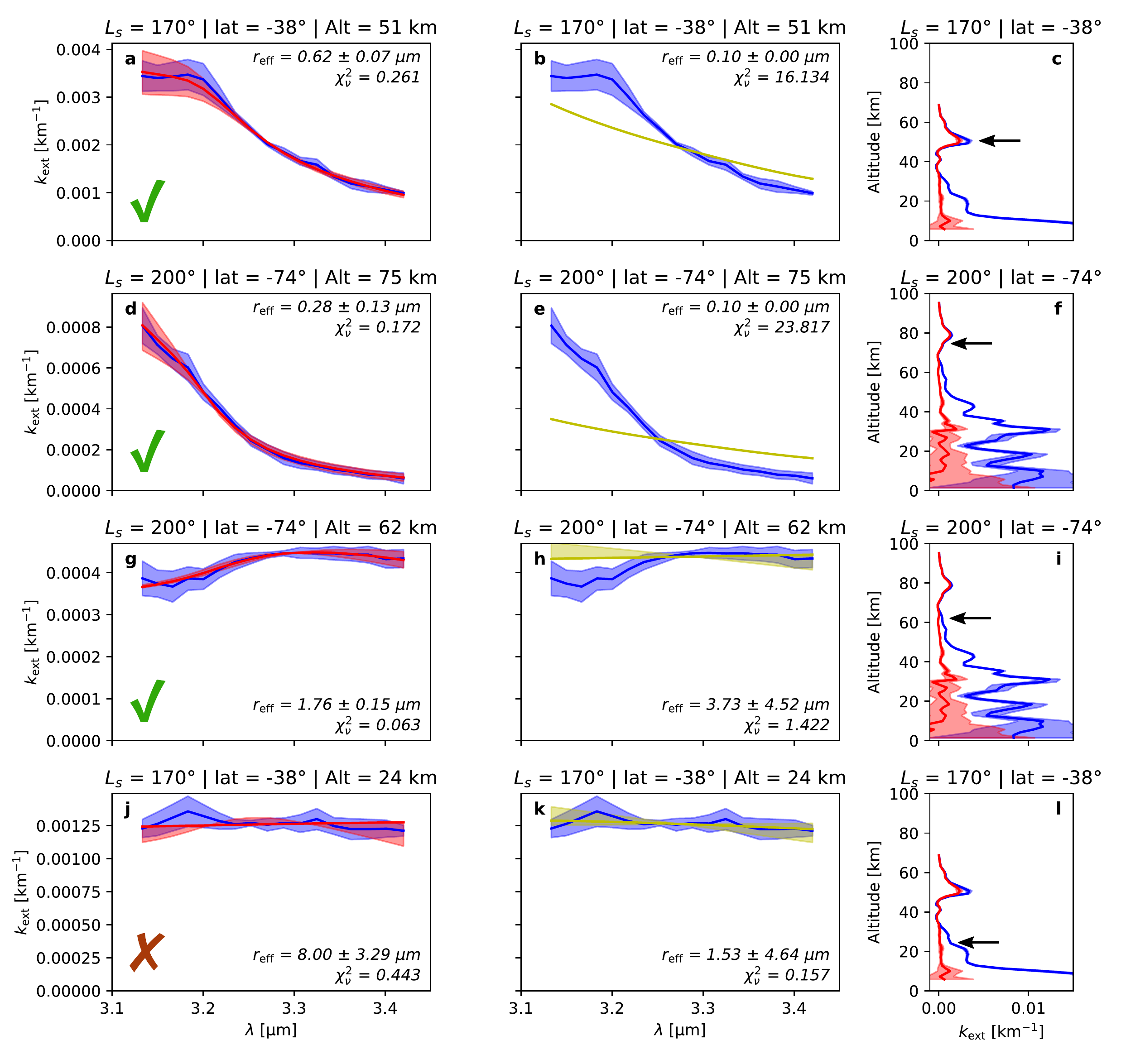}
		    \caption{Water ice cloud identification and particle size retrievals from  ACS-MIR $k_\mathrm{ext}$ spectra.
				The two first columns shows the results of the fitting algorithm using the
			    water ice model (left column) or the dust model (center column).
			    The \textcolor{blue}{blue} area represents the ACS-MIR spectra, while
			    the \textcolor{red}{red} and \textcolor{yellow!90!black}{yellow} lines
				represent respectively the best fit using the \textcolor{red}{water ice}
				and \textcolor{yellow!90!black}{dust} models.
				The right column shows for each observation (line) the associated $k_\mathrm{ext}$
				vertical profile (in \textcolor{blue}{blue} the extinction coefficient at 
				$3.2~\upmu$m, and in \textcolor{red}{red} the difference between the extinction
				coefficient at $3.2~\upmu$m and $3.4~\upmu$m).
				The black arrows indicate the altitude of the extracted spectrum.
			    We observe that even if all the presented spectra can be reproduced by our water ice
			    model, the fourth case (j) remains ambiguous as the dust model is also able to provide
			    a acceptable fit (k). 
			    Thus, we will accept the three first fits a, d \& g while the fit j will be rejected by the filtering
			    process.}
		    \label{fig:fits_reff}
		\end{figure}

		In addition, we also remove the fits whose error bars exceed a pre-defined threshold of
		0.35~$\upmu$m. An example of such a fit is found in figure~\ref{fig:fits_reff}.j.
		However, in most cases, such poorly constrained fits were already eliminated through the 
		dust-fitting test (figure~\ref{fig:fits_reff}.k).
		
		After this multi-step filtering process, we note that retrieved particle sizes never exceed 2~$\upmu$m; in essence
		the fits indicating sizes between 2~$\upmu$m and 8~$\upmu$m have been removed, which was expected as the extinction coefficient is not diagnostic of water ice in that size range, as discussed previously.
		As a consequence, it is appropriate to consider that our method makes it possible to identify and characterize water ice rich layers with mean particle size of $\sim 2~\upmu$m maximum. Other layers can correspond to either dust-dominated or large-grained water ice layers. 
		
\section{Results and discussion}    \label{sec:results}
    \subsection{The 3~$\upmu$m atmospheric absorption}    \label{sec:res_ibd}
\begin{figure}[h!]
		    \centering
		    \includegraphics[width=\textwidth]{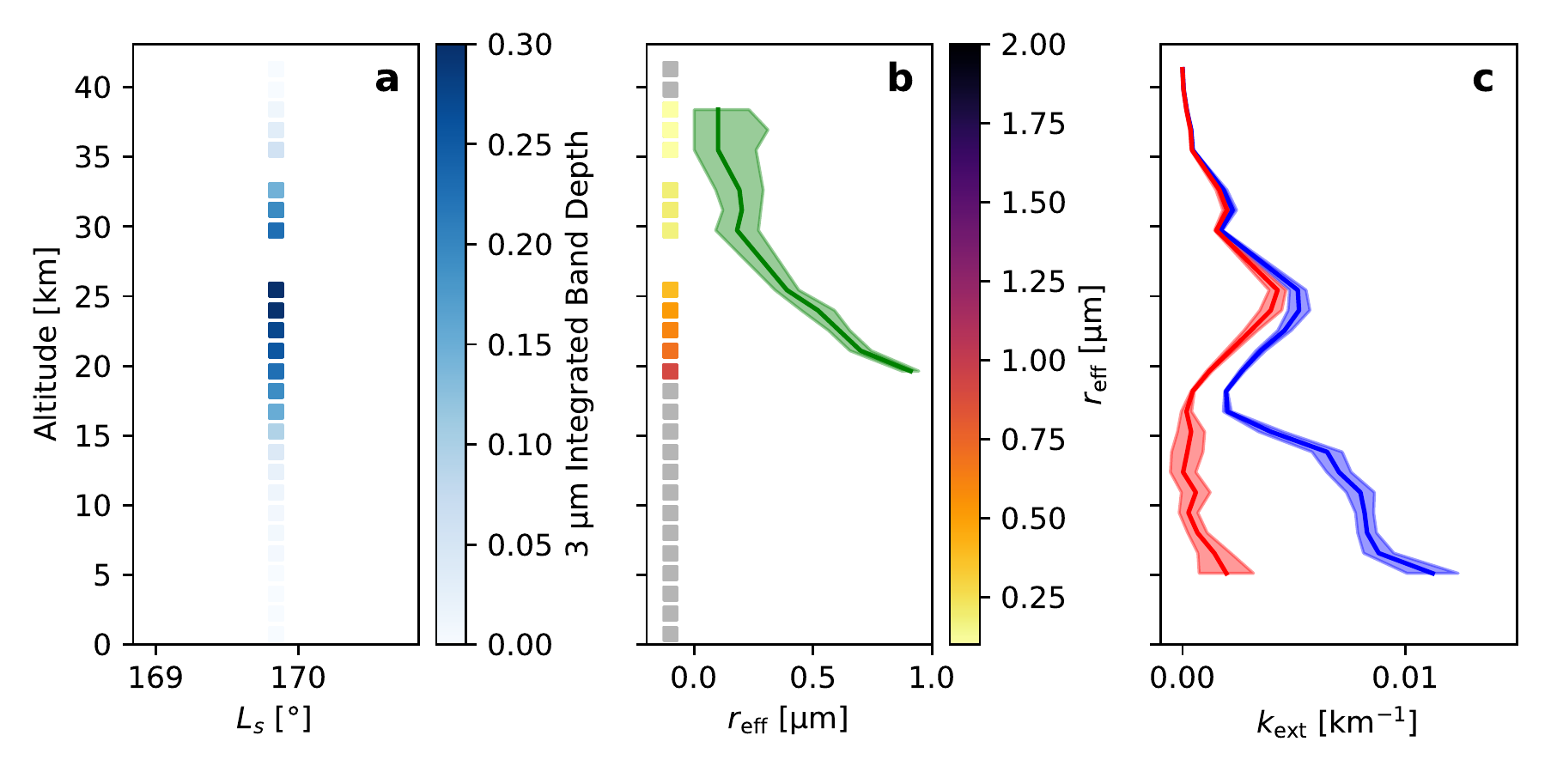}
		    \caption{Vertical profile of a water ice cloud 
		        ($L_s=169.8^\circ , \mathrm{lat} = 77^\circ \mathrm{N} , \mathrm{lon} = 152^\circ \mathrm{W}$),
		        showing the vertical variations of the IBD (\textbf{a}), particle size (\textbf{b}) and 
		        the extinction coefficient (\textbf{c}, with in \textcolor{blue}{blue} the extinction coefficient
		        at 3.2~$\upmu$m, and in \textcolor{red}{red} the difference between the extinction coefficient
		        at 3.2~$\upmu$m and 3.4~$\upmu$m).
		        We observe that the IBD criterion detects the same water ice cloud as the fitting methodology, and that
		        the size of the water ice particles of the cloud decrease as the altitude increases
		        (\textbf{b}): from 1.2~$\upmu$m at 18~km to 0.2~$\upmu$m at 31~km.}
		    \label{fig:profil_ibd_reff}
		\end{figure}

\begin{figure}[h!]
			\centering
			\includegraphics[width=\textwidth]{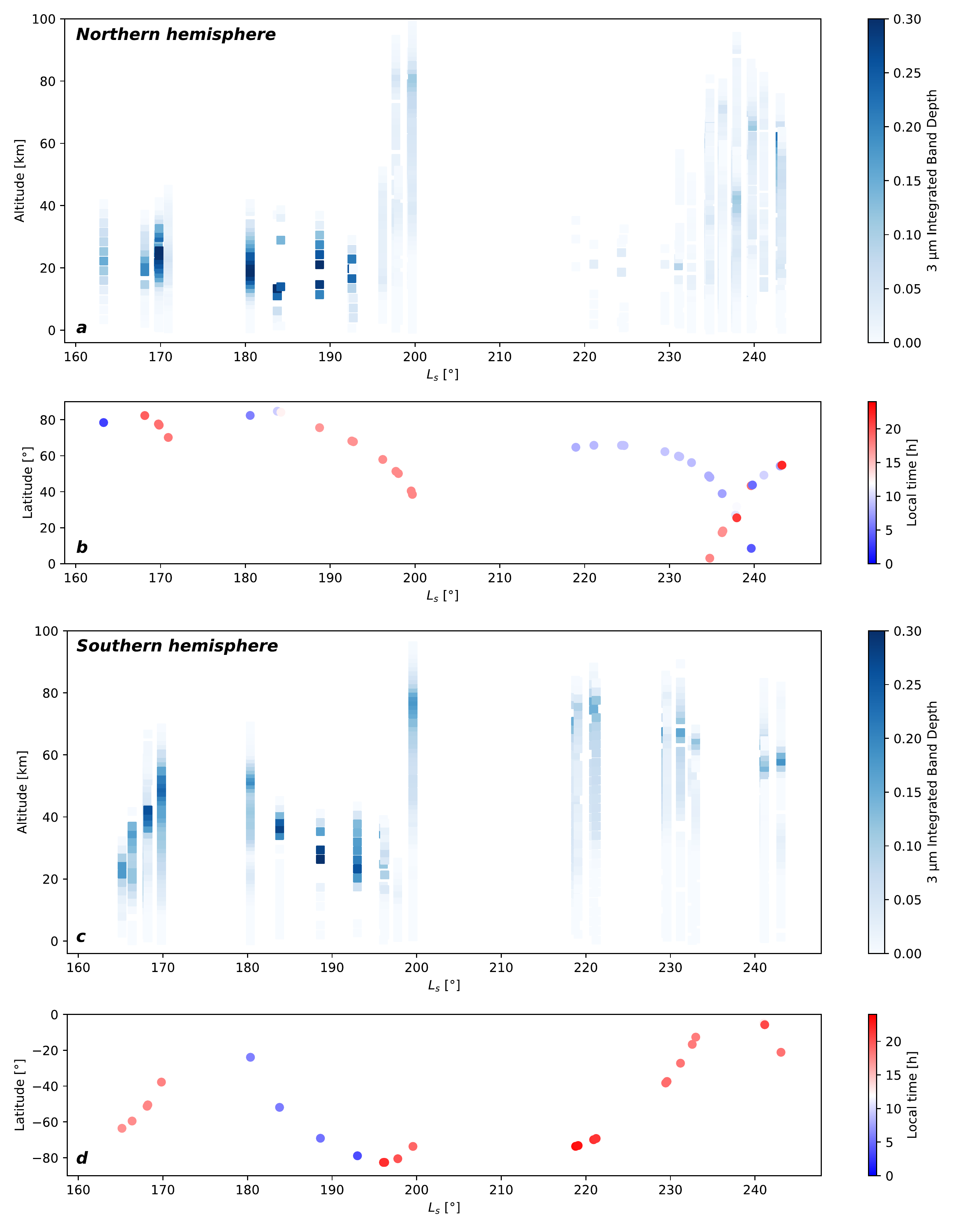}
			\caption{3~$\upmu$m water ice absorption band (IBD) monitoring from the ACS MIR channel in the
					Martian Northern (\textbf{a}) and Southern (\textbf{c}) hemisphere, before and during the 2018 GDS (MY 34). 
					Latitude and local time of observations are indicated in panels \textbf{b}~\&~\textbf{d}}.
			\label{fig:IBD_profils}
		\end{figure}

        The atmospheric 3~$\upmu$m integrated band depth can be used to obtain a quick look at water ice clouds in our dataset.
        It is also of interest in itself as orbital observations of Mars surface have to account for atmospheric 3~$\upmu$m contribution to deliver information about surface hydration \protect\cite{jouglet_2007, audouard_2014}.
        As illustrated in figure~\ref{fig:profil_ibd_reff}, the 3~$\upmu$m IBD matches at first order the more elaborated water ice detection scheme presented in section~\ref{sec:clouds_prop}. 
	    Actually, the 3~$\upmu$m IBD captures only small-grained ($r_\mathrm{eff} \leq 1~\upmu$m) water ice
	    clouds according to modeling results presented in figure~\ref{fig:models_Mike}, but this
	    corresponds to typically 90\% of detected clouds here 
	    (see figures~{\protect\ref{fig:reff_alt}}~\&~{\protect\ref{fig:distrib_reff}}).
	    In addition, the IBD does not capture thin clouds that produce a tenuous absorption, whatever their grain size. 
	    Thus, thin layers (normal integrated layer optical depth lower than 0.01) that can be observed at very high altitudes ($\geq$ 90~km) are not always discernible with IBD figures.
	    Beside these slight differences, the IBD criteria thus illustrates the overall water ice pattern and 
	    provide a proxy for the optical thickness of clouds. We have illustrated the overall behavior 
	    of water ice clouds using figure~\ref{fig:IBD_profils} where the 3~$\upmu$m IBD variations 
	    are represented from $L_s~=~165^\circ$ to $L_s~=~243^\circ$ in the Northern (a) and Southern (c) hemispheres.
	    
	    We observe in figure~{\protect\ref{fig:IBD_profils}} that some vertical profiles corresponding to
	    observations acquired at close $L_s$ and are so overlapping in the figure
	    can manifest notable differences in the 3~$\upmu$m IBD criterion. 
	    This probably reflects actual variations between profiles, associated with longitudinal variations 
		of the water ice clouds \protect\cite{smith_2004, wolff_2019}, as illustrated in table~\ref{tab:overlaps}. 
	    Local time differences \cite{szantai_2019} may also contribute to the variability
	    (Table~\ref{tab:overlaps}), although the dataset is not yet extensive enough
	    to properly delineate local time effects. 
	
		\begin{table}[h!]
		    \centering
		    \begin{tabular}{ccccl}
		        $L_s$ & Latitude & Longitude & Local time & \textit{Observation} \\
		        \hline
		        \hline
		        $168.1^\circ$ & $51^\circ$S & \textbf{163$^\circ$W} & 17h39 & \textit{No IBD} \\
		        $168.2^\circ$ & $50^\circ$S & \textbf{139$^\circ$E} & 17h40 & \textit{IBD $\sim$ 0.25 at 40 km} \\
		        \hline
		        $192.6^\circ$ & $68^\circ$N & \textbf{28$^\circ$E}  & 17h09 & \textit{IBD $\sim$ 0.26 at 20 km} \\
		        $192.7^\circ$ & $68^\circ$N & \textbf{ 3$^\circ$E}  & 17h09 & \textit{No IBD} \\
		        \hline
		        $229.4^\circ$ & $38^\circ$S & \textbf{136$^\circ$E} & 18h53 & \textit{IBD $\sim$ 0.18 at 65 km} \\
		        $229.6^\circ$ & $37^\circ$S & \textbf{ 50$^\circ$E} & 18h52 & \textit{No IBD} \\
		        \hline
		        $243.1^\circ$ & $54^\circ$N & \textbf{70$^\circ$E}  & \textbf{07h50} & \textit{IBD $\sim$ 0.22 at 60 km} \\
		        $243.3^\circ$ & $55^\circ$N & \textbf{45$^\circ$W}  & \textbf{22h10} & \textit{No IBD} \\
		        \hline
		    \end{tabular}
		    \caption{Identification of consecutive observations with and without 
		        3~$\upmu$m atmospheric absorption using the IBD, showing the local aspect of our observed clouds
		        (see the vertical IBD profiles in figure~\ref{fig:IBD_profils}). 
		        The first three examples are associated with significant variations in longitude, 
				while the last example also show strong local time variations.}
		    \label{tab:overlaps}
		\end{table}
    
	\subsection{Clouds monitoring}  \label{sec:clouds_monitoring}
\begin{figure}[h!]
            \centering
            \includegraphics[width=\textwidth]{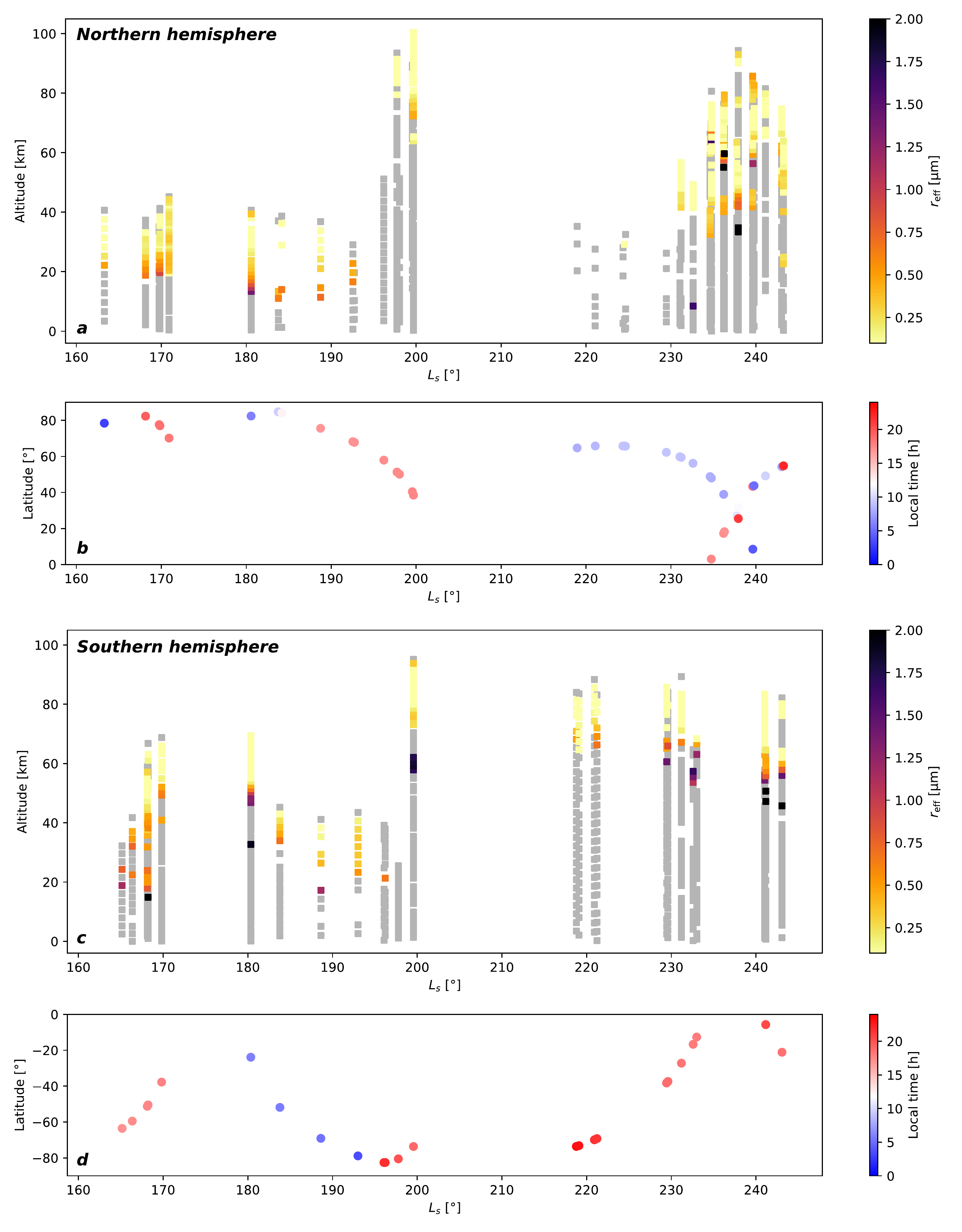}
            \caption{Vertical profiles of the retrieved size of the water ice particles in the 
                    Martian Northern (\textbf{a}) and Southern (\textbf{c}) hemispheres, before and during the MY 34 GDS.
                    In \textcolor{gray}{grey} the observations without water ice detection.
                    Latitude and local time of the observations are indicated in panels \textbf{b}~\&~\textbf{d}.}
            \label{fig:profils_reff}
        \end{figure}

	    Using the method presented in section~\ref{sec:clouds_prop}, we can both detect the
	    presence of water ice clouds layers in the Martian atmosphere and retrieve their particle size
	    (see discussion in section~\ref{sec:res_part_size}).
	    Figure~\ref{fig:profils_reff} presents the resulting vertical profiles of water ice particles
	    from $L_s~=~165^\circ$ to $L_s~=~243^\circ$ in the Northern and Southern hemispheres.
		This overview reveals two main trends in our dataset. First, we detect water ice clouds in most observations. 
		In more detail, in our dataset of 65 ACS-MIR grating position-12 observations, 11 of them have no 
		water ice cloud detection, i.e., 83\% of our profiles show at least one detection of water ice. Non-detections are localized in time and places: 45\% of our non-detections are located at the onset of the GDS ($193^\circ~\leq~L_s~\leq~198^\circ$, 5 observations), and the other 55\% are located during the GDS ($219^\circ~\leq~L_s~\leq~231^\circ$, 6 observations) but corresponds to low haze top altitude beyond the northern limit of the GDS (see detailed discussion in section~\ref{sec:during_gds}). 
		Second, we observe the presence of two distinct types of vertical profiles that correspond to the two periods separated by the sudden onset of the GDS at $L_s~\sim~195^\circ-200^\circ$ \cite{guzewich_GDS_2019}. 
		Two main cloud characteristics differ between these two periods: the vertical extent, and the water ice cloud opacity. Most of the detected clouds are between 10~km and 70~km before the onset of the GDS, while clouds altitudes typically extend higher between 60~km and about 90~km afterward. 
		Figure~\ref{fig:profils_kext} shows the vertical profiles of $k_\mathrm{ext}$ at
		3.4~$\upmu$m (i.e. the opacity per km). 
		As 3.4~$\upmu$m is on the edge of the 3~$\upmu$m water ice band \cite{vincendon_2011}
		it provides an estimator of the global atmospheric opacity,
		whereas the opacity at 3.2~$\upmu$m is strongly increased by the presence of water ice clouds.
		We observe that the water ice clouds opacity at 3.2~$\upmu$m (vertically integrated along the cloud) typically goes from 0.01 to 0.05 before the GDS, but falls to
	    a few $10^{-3}$ for the mesospheric clouds during the GDS, with $k_\mathrm{ext}(\lambda=3.2~\upmu\mathrm{m})\sim 10^{-4}~\mathrm{km}^{-1}$ around 90~km
	    (and $k_\mathrm{ext}(\lambda=3.4~\upmu\mathrm{m}) \sim 10^{-5}~\mathrm{km}^{-1}$). 
	    We discuss in more details the distribution and properties of clouds during these two periods in the next two sections.

		\begin{figure}[h!]
		    \centering
		    \includegraphics[width=\textwidth]{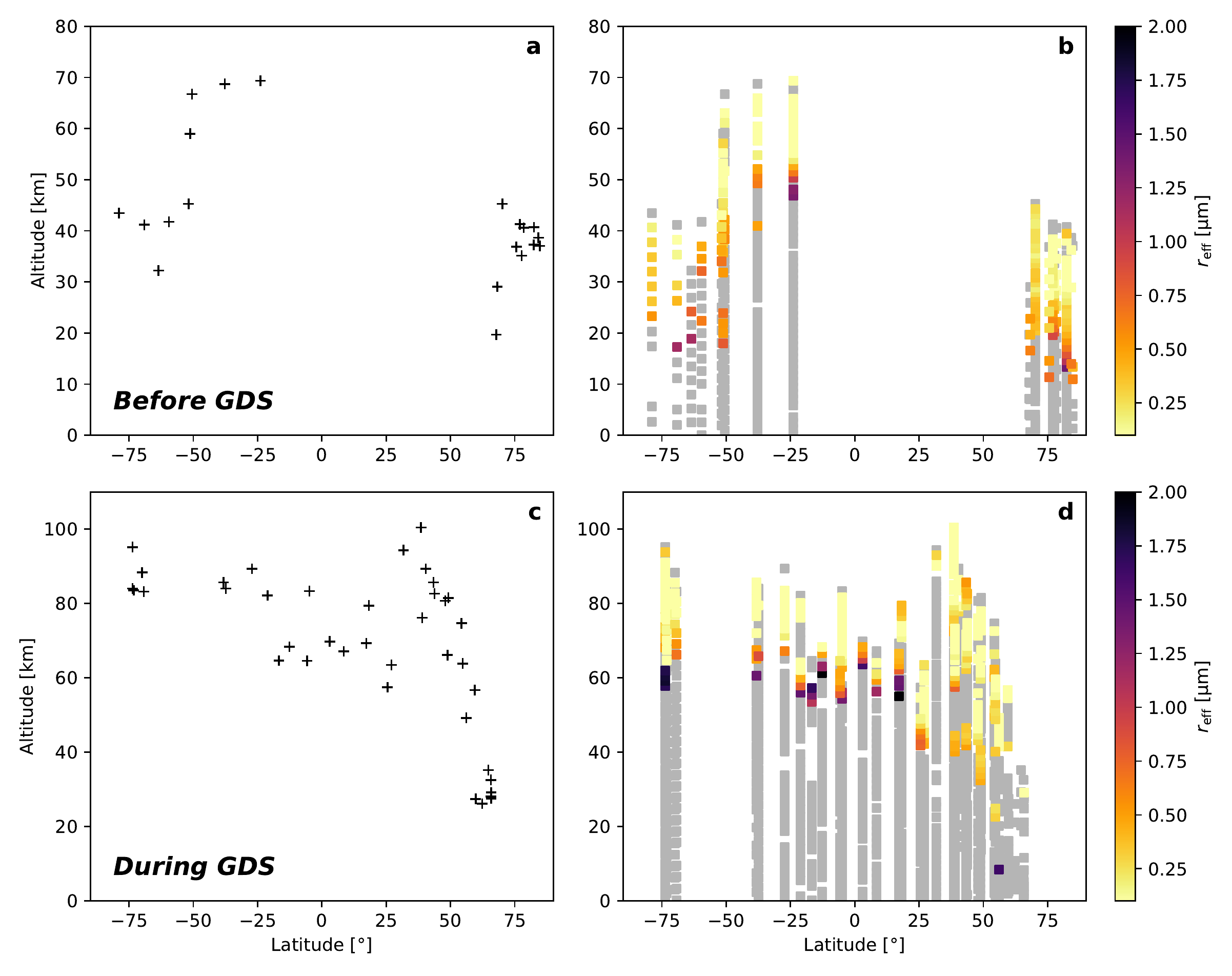}
		    \caption{\textbf{a.} The altitude of the haze top as a function of latitude 
		        for the observations before the MY 34 GDS
		        ($163^\circ \leq L_s \leq 195^\circ$).
		        We observe a latitude dependence of the haze top which increases equatorwards
		        in the Southern hemisphere. Unfortunately, all the observations of the Northern hemisphere
		        during these period are concentrated at high latitudes.
		        \textbf{b.} The vertical profiles of $r_\mathrm{eff}$ as a function of latitude 
		        for the same observations.
		        In the Southern hemisphere, we observe that the water ice clouds altitude
		        seems to follow the same latitudinal trend of the haze top in the
		        equatorial region.\\
		        \textbf{c. \& d.} Same as a. \& b. but during the MY 34 GDS
		        ($199^\circ\leq L_s \leq 244^\circ$).}
		    \label{fig:htop_profils}
		\end{figure}
		
        \begin{figure}[h!]
            \centering
            \includegraphics[width=\textwidth]{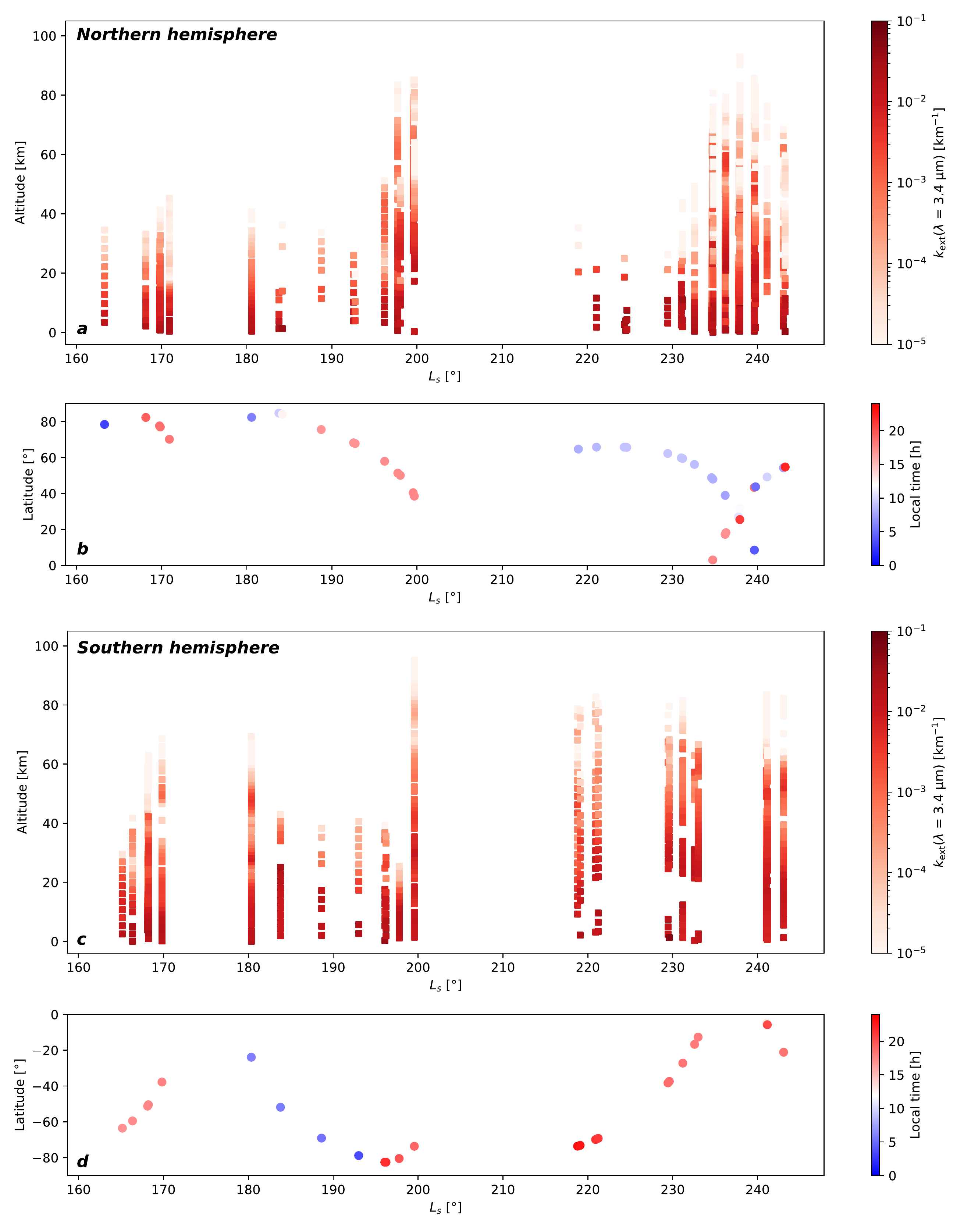}
            \caption{Vertical profiles of the measured opacity 
                    ($k_\mathrm{ext}=\mathrm{d}\tau/\mathrm{d}z$) at 3.4~$\upmu$m in the 
                    Martian Northern (\textbf{a}) and Southern (\textbf{c}) hemispheres, before and during the MY 34 GDS.
                    Latitude and local time of the observations are indicated in panels \textbf{b}~\&~\textbf{d}.}
            \label{fig:profils_kext}
        \end{figure}

		\subsection{Before the GDS ($L_s~<~195^\circ$)}	\label{sec:before_gds}
			Before the dust storm, we observe the presence of water ice clouds in most profiles, with altitudes ranging from 11 
			to 44~km in the Northern hemisphere, and between 17 and 69~km in the Southern hemisphere, 
			which is consistent with previous observations \cite{smith_2013}.
			Thus, clouds are found below the mesosphere for most of the cases. However, the latitude coverage of available 
			observations in the North and in the South are not equivalent (cf. figure~\ref{fig:profils_reff}): 
			Northern observations prior to the GDS are restricted to polar latitudes $\geq 68^\circ$.
			
			During this period, we indeed observe a latitude-dependence of the haze top altitude that increases 
			closer to the equator (up to $\sim$70~km at 24$^\circ$S) and decreases when moving towards 
			polar latitudes (down to $\sim$40~km at 80$^\circ$N, or 60$^\circ$S), see also 
			figure~\ref{fig:htop_profils}.a. This trend is consistent with the expected haze top behavior
			\cite{jaquin_1986, forget_1999, montmessin_2006, heavens_2011a, smith_2013}. 
			However, the lack of data around the equatorial latitudes before the GDS (no observations
			between 20$^\circ$S and 60$^\circ$N, see figure~\ref{fig:htop_profils}.a) prevents us to
			get access to the latitude of the maximum haze top altitude.
			
			Water ice clouds are detected in the upper parts of most profiles, but some profiles also include
			lower detached layers (e.g. figure~\ref{fig:kext_vert}.f at 30~km) that are not detected either by 
			the fitting process or the IBD (as the $k_\mathrm{ext}$ at 3.2~$\upmu$m and 3.4~$\upmu$m are similar).
			Thus, these layers are likely be either composed primarily of dust or by $>2~\upmu$m water ice particles 
			(cf. section~\ref{sec:filtering}). 
			This suggests that water ice appears to frequently cap the dust layer in our solar longitude range \cite{smith_2013}. 
			Note however that larger-grained water ice clouds can also be present at lower altitude \cite{wolff_2003}, 
			but remain undetected here due to a lack of a clear signature in our spectral range (as discussed previously). 
			Nevertheless, the altitude distribution of the detected water ice clouds follows a similar latitudinal trend as the haze top itself:
			the average altitude of cloud detection prior the dust storm increases equatorward (cf. figure~\ref{fig:htop_profils}.b),
			as previously noticed by \citeA{kleinbohl_2009} and \citeA{smith_2013}. 

		\subsection{During the GDS ($L_s~>~200^\circ$)}	\label{sec:during_gds}
			Around $L_s~=~200^\circ$, there is a sudden increase in the altitudes of both 
			the haze top (up to 100~km) and the  water ice clouds ($\geq$ 90~km), while geographical coordinates remain essentially 
			the same (83$^\circ$S and 58$^\circ$N at $L_s~=~196^\circ$, and
			74$^\circ$S and 39$^\circ$N at $L_s~=~200^\circ$).
			These behaviors are combined with a decrease of the measured IBD values (cf. figure~\ref{fig:IBD_profils}).
			We note that the above behavior appears to be relatively uniform around 
			the planet (i.e., both zonally and meridionally).
			Specifically, in the Southern hemisphere, even though the observed latitude varies from 74$^\circ$S to 5$^\circ$S
			between $L_s~=~219^\circ$ and $L_s~=~241^\circ$, one does not see a change in both
			the haze top altitude and the maximal altitude of the water ice clouds; in agreement with
			the simulations presented in \citeA[Figure 3]{neary_2019}.
			We nonetheless identify a temporal trend likely related to the progressive decay of the dust storm \cite{guzewich_GDS_2019}:
			from $L_s=220^\circ$ to $L_s=243^\circ$ the lower altitude of the detected water ice clouds
			in the Southern hemisphere decrease 
			from 68~km to 54~km (cf. figure~\ref{fig:profils_reff}) 
			and the main 3~$\upmu$m atmospheric absorption (i.e. the maximum of IBD for a given profile)
			decrease from 75~km to 58~km (cf. figure~{\protect\ref{fig:IBD_profils}}).
			Regardless that the haze top altitude remains at $\sim$ 85 -- 90~km, 
			and the latitude goes successively towards and from the equator. 
			Nevertheless, we note a slight North/South asymmetry during the dust storm that is represented by the smaller number of high-altitude water ice clouds in the Northern hemisphere (see figure~\ref{fig:htop_profils}.c\&d).

			Ice clouds are detected at very high altitudes in the mesosphere during this period, up to $\geq$ 80~km at the 
			beginning of the GDS. A few profiles show robust water detections at altitudes above 90~km, 
			up to 100~km maximum at $L_s~\sim~200^\circ$ (cf. figure~\ref{fig:profils_typiques_reff}.c). 
			These high altitude water ice clouds are observed only at the beginning of the GDS. 
			High altitude water ice clouds can also be seen during non-dust GDS years, 
			when they occur during the storm season \cite{vincendon_2011, clancy_2019}. 
			However, high cloud altitudes are nearly systematic in our observations during the peak phase of the GDS 
			(cf. figure~\ref{fig:profils_reff}), while such altitudes are less typical during non-GDS years \cite{clancy_2019}. 
			Moreover, peak cloud altitudes are slightly higher during the onset of the GDS than reported in non-GDS years. 
			Indeed, models \cite{neary_2019} showed that GDS are expected to slightly increase the average and maximum altitude 
			of water ice clouds, which is consistent with our observations. 
    		The sharp increase in water ice cloud altitude observed during the GDS indeed confirms that this water ice cloud 
			increase is directly connected to dust storm activity. The large-scale dust storm probably facilitates the formation 
			of high altitude water ice clouds through increase in the altitude of both water vapor and condensation nuclei. 
			This increase of water ice altitude during the GDS is indeed consistent with the recently reported increase 
			of water vapor at higher altitudes \cite{fedorova_2018, neary_2019}. 
    		
    		We still observe a North/South latitudinal asymmetry during the GDS. 
			In the Northern hemisphere, seven observations in the figure~\ref{fig:htop_profils}.c
    		exhibit low haze top altitudes during the GDS. The haze top is between 26~km and 
    		36~km, i.e. $\sim$~50~km below the altitudes measured during the dust storm for most of the other profiles.
    		Actually, these profiles were taken at high Northern latitude (60$^\circ$N -- 66$^\circ$N)
    		during the decay of the dust storm ($L_s = 221^\circ - 231^\circ$) at
    		longitudes of $\sim25^\circ$W (Acidalia Planitia) and $\sim155^\circ$E (Utopia Planitia),
    		see table~\ref{tab:pts_low_htop}: this suggests that 60$^\circ$N represents the Northern extent of the dust storm activity
			in these regions from $L_s\sim220^\circ$, which is in agreement with MCS measurements
			\cite{kass_2019}.
			The scheme is different for the Southern hemisphere, with high haze top
			altitudes ($> 80$~km) observed during the GDS up to the polar regions ($73^\circ$S).

    		\begin{table}[h!]
    		    \centering
    		    \begin{tabular}{ccccc}
    		        $L_s$ & Latitude & Longitude & Local time & Haze top altitude \\
    		        \hline
    		        \hline
    		        $219.0^\circ$ & $65^\circ$N & 113$^\circ$E & 08h14 & 35.3 km \\
    		        $221.1^\circ$ & $66^\circ$N &   6$^\circ$W & 08h38 & 27.5 km \\
    		        $224.3^\circ$ & $66^\circ$N &  38$^\circ$W & 09h07 & 28.1 km \\
    		        $224.6^\circ$ & $66^\circ$N & 152$^\circ$W & 09h09 & 29.2 km \\
    		        $224.7^\circ$ & $66^\circ$N & 150$^\circ$E & 09h09 & 32.5 km \\
    		        $229.5^\circ$ & $62^\circ$N &  28$^\circ$W & 09h16 & 26.2 km \\
    		        $231.1^\circ$ & $60^\circ$N & 159$^\circ$E & 09h06 & 27.4 km \\
    		        \hline
    		    \end{tabular}
    		    \caption{Properties of profiles located outside the GDS beyond its Northern limit 
                    at $L_s\sim225^\circ$. 
    		        With a mean haze top altitude of 30~km, i.e. 50~km lower other measurements during the same
    		        period, these observations can no longer be considered as part of the GDS.
                    These low haze top observations are located at latitudes greater than
    		        $60^\circ$ with some longitudinal variability. This latitude thus correspond to the Northern 
    		        maximum extent of the GDS at that time, in agreement with MCS measurements \cite{kass_2019}.}
    		    \label{tab:pts_low_htop}
    		\end{table}

	\subsection{Particle size}  \label{sec:res_part_size}
\begin{figure}[h!]
            \centering
            \includegraphics[width=\textwidth]{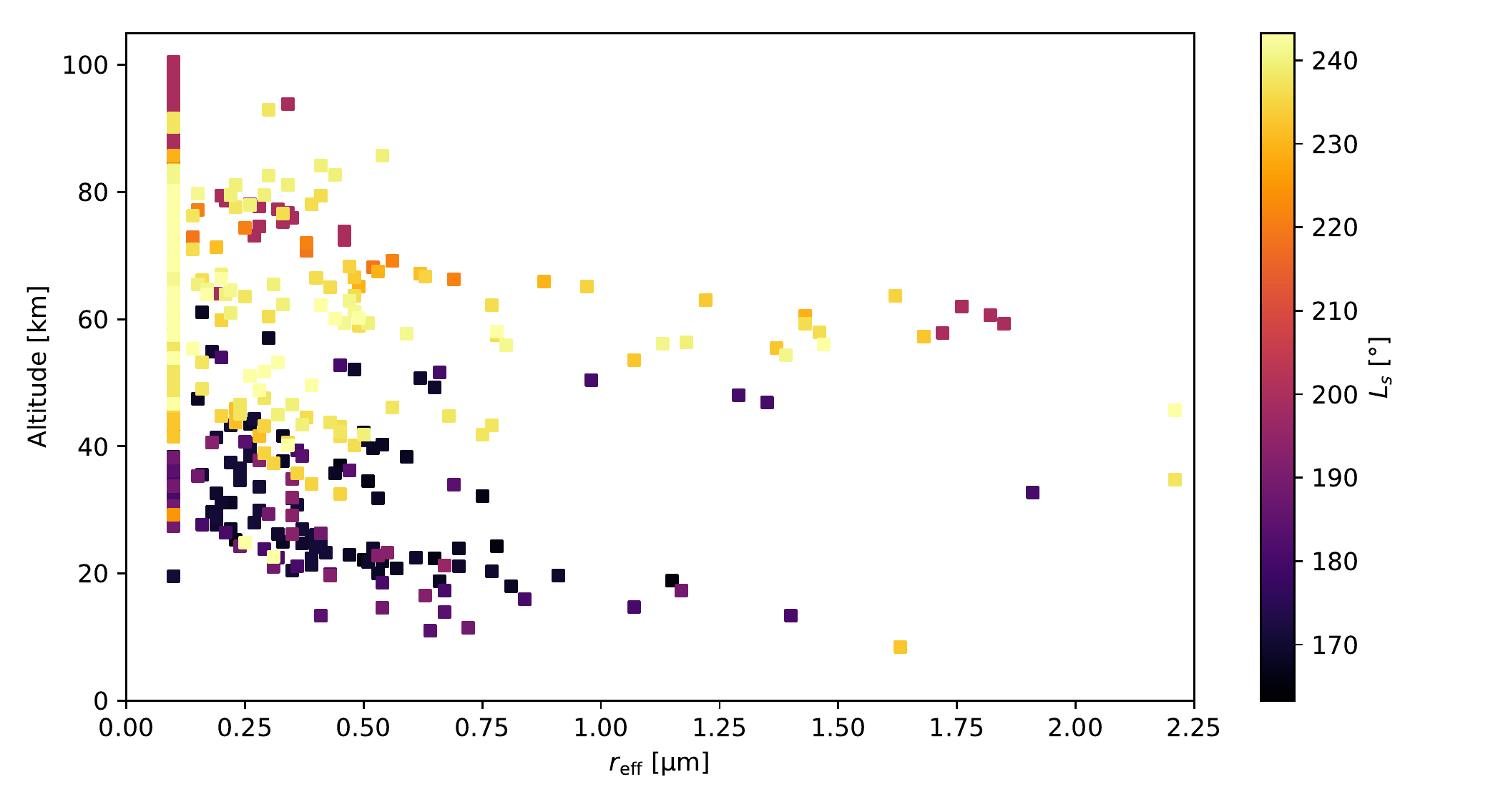}
            \caption{Distribution of the water ice particles size, as a function of the observed
                    altitude and the Solar longitude.
                    During the dust storm ($L_s\geq200^\circ$) the altitude of the water ice
                    clouds increase, and they are mostly composed by smaller particles than
                    before the dust storm (see also figure~\ref{fig:distrib_reff}).}
            \label{fig:reff_alt}
        \end{figure}

\begin{figure}[h!]
		    \centering
		    \includegraphics[width=\textwidth]{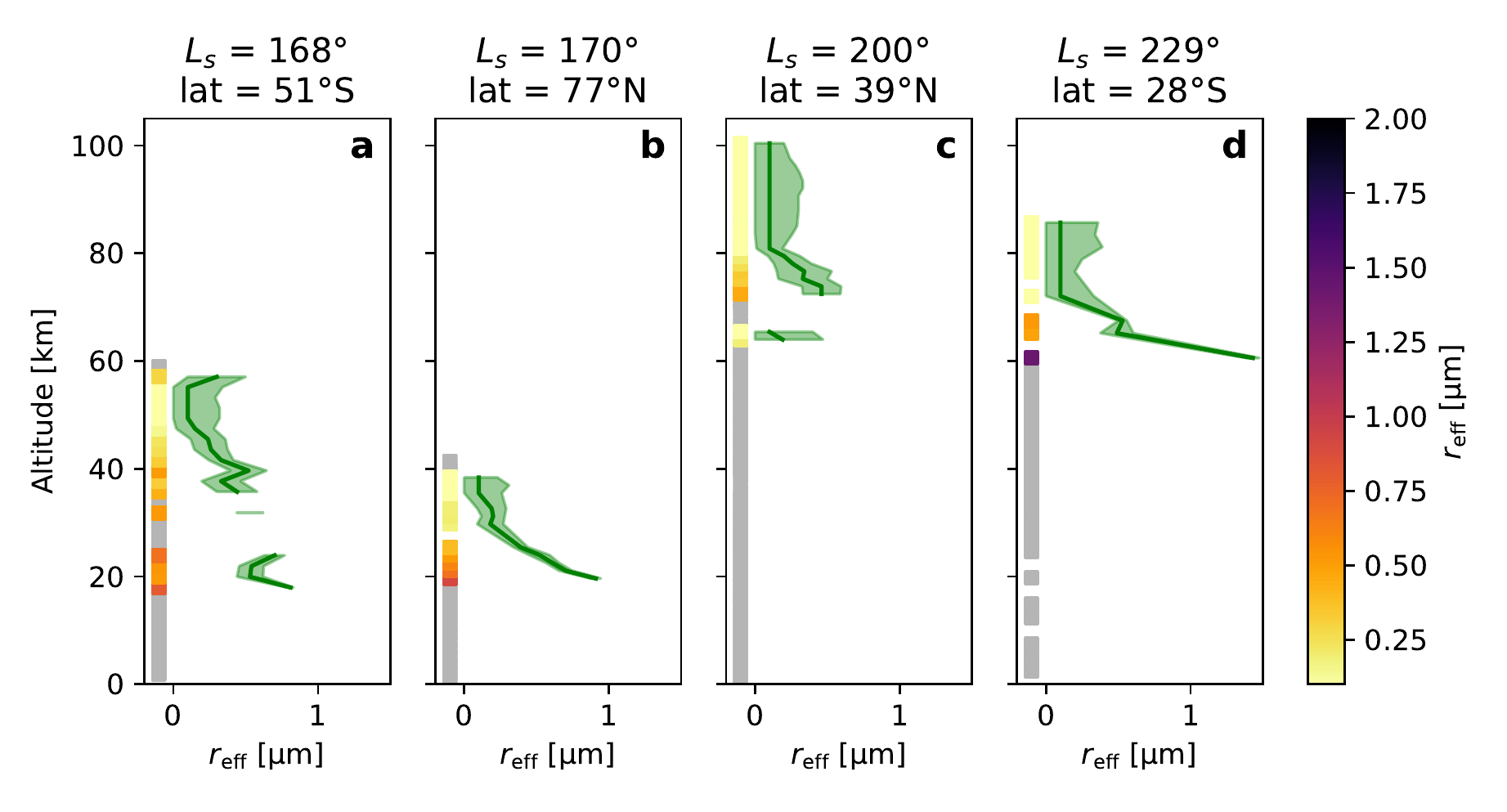}
		    \caption{Typical vertical profiles of water ice clouds particle size before (\textbf{a}~\&~\textbf{b})
		        and during (\textbf{c}~\&~\textbf{d}) the GDS.
		        Each panel represents the same profile in two manners: first the vertical line
		        on the left where $r_\mathrm{eff}$ is represented using a color scale (same as
		        figure~\ref{fig:profils_reff}); 
		        and second the \textcolor{green!50!black}{green} lines with $r_\mathrm{eff}$
		        on the x-axis and the uncertainties shown by the shadowed regions.
		        We observe that the size of the water ice particles of the cloud decreases as the altitude 
		        increases, regardless of the mean altitude of the cloud.
		        The GDS is characterized by a shift in altitude of the particle size distribution
		        ($\sim$~30~km higher, see profile~\textbf{c} compared to \textbf{a} and \textbf{b}), 
		        along with a decrease of particles with size
		        between 0.5~$\upmu$m and 1~$\upmu$m (cf. figure~\ref{fig:distrib_reff}).
		        But some GDS profiles show evidence of larger-grained layers ($> 1~\upmu$m) as we can see 
		        at 60~km in the profile~\textbf{d}.}
		    \label{fig:profils_typiques_reff}
		\end{figure}

		We observe that the particle size decreases on average with increasing altitude (cf. figure~\ref{fig:reff_alt}),
		as previously reported by \citeA{clancy_2019}.
		This global trend is also apparent in the majority of individual profiles: particle size is observed to decrease with 
		altitude within a given cloud as we can see in figure~\ref{fig:profils_typiques_reff}. 
		This is true regardless of the average altitude of the considered cloud (cf. figure~\ref{fig:profils_typiques_reff}).
		The distribution of particle size as a function of altitude is globally shifted to higher altitudes during the GDS 
		(cf. figure~\ref{fig:reff_alt}), i.e., a given size of particles can go higher in the atmosphere during the GDS.
		\citeA{clancy_2019} also observe a similar trend during non-GDS years, with a given class of particle size 
		that increases in altitude during the perihelion season ($L_s~=~180-340^\circ$).
		
\begin{figure}[h!]
            \centering
            \includegraphics[width=\textwidth]{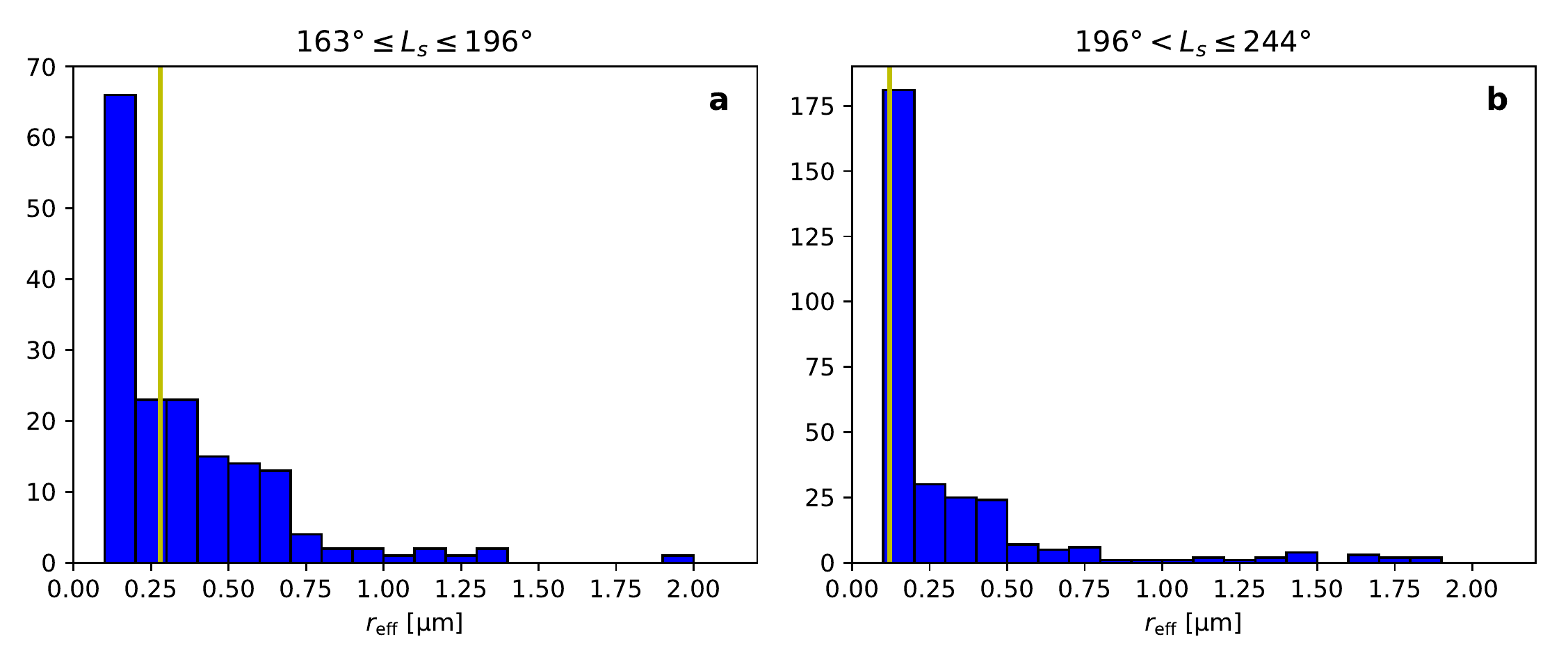}
            \caption{Distribution of the retrieved water ice particles sizes, before (a) and 
				during (b) the MY 34 GDS. 
				The \textcolor{yellow!90!black}{yellow} lines represent the median size for each 
				distribution, respectively $0.28~\upmu$m (a) and $0.1~\upmu$m (b).
                We observe that during the dust storm we detect a fewer proportion of particle with
                a size between $0.5~\upmu$m and $1~\upmu$m, as most of our detection suggest sizes
                lower than $0.5~\upmu$m, with a strong peak at $0.1~\upmu$m.
                However, because we do not derive particles sizes lower than $0.1~\upmu$m, 
                the median $r_\mathrm{eff}$ value during the GDS must be consider as
                $\leq 0.1~\upmu$m as more than half of our retrievals indicates sizes on the lower bounds
                of the models.}
            \label{fig:distrib_reff}
        \end{figure}

		While trends can be identified on a global scale, some variability is also evident, as seen in 
		figure~\ref{fig:distrib_reff}. As the main clouds altitude increases, water ice particles become distinctly smaller in the storm, 
		with a median effective radius of $0.28~\upmu$m before the dust storm and $\leq$ $0.1~\upmu$m during it.
		This narrower range of smaller particles, with a strong peak for $r_\mathrm{eff}\leq$0.2~$\upmu$m,
		is typical of the mesospheric clouds \cite[figure~12]{clancy_2019}.
		
\begin{figure}[h!]
			\centering
			\includegraphics[width=\textwidth]{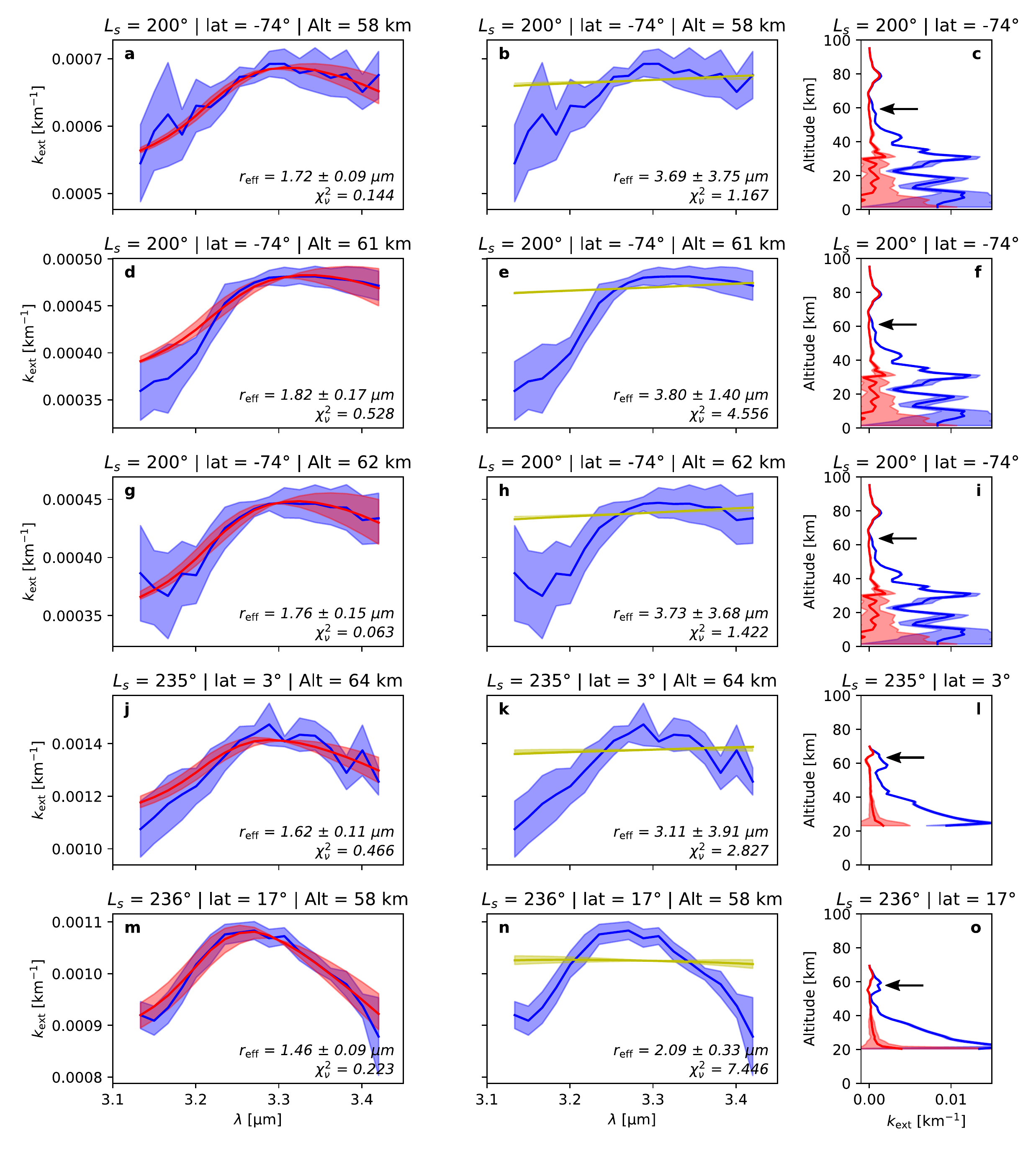}
			\caption{Big water ice particles ($r_\mathrm{eff}~\geq~1.5~\upmu$m) during the GDS.
				The two first columns shows the results of the fitting algorithm using the
			    water ice model (left column) or the dust model (center column).
			    The \textcolor{blue}{blue} area represents the ACS-MIR spectra, while
			    the \textcolor{red}{red} and \textcolor{yellow!90!black}{yellow} lines
				represent respectively the best fit using the \textcolor{red}{water ice}
				and \textcolor{yellow!90!black}{dust} models.
				The right column shows for each observation (line) the associated $k_\mathrm{ext}$
				vertical profile (in \textcolor{blue}{blue} the extinction coefficient at 
				$3.2~\upmu$m, and in \textcolor{red}{red} the difference between the extinction
				coefficient at $3.2~\upmu$m and $3.4~\upmu$m).
				The black arrows indicate the altitude of the big particle observation.
				We can see in this figure that the detection of these large water ice particles is robust,
				as the full-dust model is not able to reproduce the observed spectra, especially the
				bump of extinction coefficient with the wavelength.}
			\label{fig:fits_gros_grains}
		\end{figure}

		Additionally, we also observe the appearance of a small population of larger water
		ice particles, with $r_\mathrm{eff}\geq1.5~\upmu$m, at surprisingly high altitude
		(between 55~km and 64~km) in view of previous observations and studies
		\cite<e.g. below 40 -- 50~km for>{guzewich_2019}.
		Their identification as water ice particles is robust, i.e., attempts to fit using dust properties
	    are unable to reproduce the observed wavelength dependence of the derived $k_\mathrm{ext}$
		(cf. figure~\ref{fig:fits_gros_grains}).
		However, we have to keep in mind that our understanding of the spectral behaviour of the
		martian dust still needs to be improved, in particular about the hydration properties.
		Nevertheless, even updated dust optical properties are not expected to be able to reproduce the spectral bump observed
		in figure~\ref{fig:fits_gros_grains}, as large water ice particles can do 
		(cf. figure~\ref{fig:models_Mike}).
		These detections are found below smaller water ice particles (cf. figure~\ref{fig:fits_gros_grains}, 3rd column).
		Although turbulence during dust storms is expected to enable the lifting of larger particles, the exact mechanism 
		by which these large-grained water ice clouds form or are bring at high altitude remains to be investigated. 
		In addition to turbulence, condensation nuclei properties could also play a role 
		\cite{gooding_1986, michelangeli_1993, montmessin_2004, hartwick_2019}. 
		One possibility is that these particles are not primarily composed of water ice but could host a relatively large dusty core. 
		Incidentally, large dust particles have been reported from the surface during this GDS, 
		\cite{lemmon_2019}, which could play a role as condensation nuclei for the water ice. 
		\citeA{clancy_2010} indeed observed during a previous GDS (2001) that large dust particles can propagate to high
		altitude (sizes between 1~$\upmu$m and 2~$\upmu$m above 70~km).
    		
\section{Conclusion}	\label{sec:ccl}
    In this paper, we present our analyses of the Martian water ice clouds in the period before 
	and during the MY 34 GDS using ACS infrared SO observations. This novel 
	observational geometry in the mid-infrared allows us to monitor the 3~$\upmu$m water ice 
	absorption band, and subsequently constrain the sizes of the detected water ice particles. 
    Indeed, the shape of the wavelength profile of the 3~$\upmu$m absorption band is a 
	sensitive function of the particle size. The sensitivity of this feature combined with the 
	high resolution of ACS-MIR allows us to derive useful constraints on the particle size 
	distribution of the clouds between $r_\mathrm{eff}~\leq~0.1~\upmu$m and 2~$\upmu$m.
    Moreover, the temporal range of this dataset, from $L_s=165^\circ$ to 
	$L_s=243^\circ$, offers an unique opportunity to observe the impact of such a dust storm on the 
	distribution and composition of the water ice clouds in the Martian atmosphere.
    
    The main results are summarized below:
    \begin{enumerate}
        \item Using the ACS MIR dataset, we have developed of a method to identify water ice clouds, 
            measure their opacity, and estimate their particle size. 
            The method makes it possible to identify cloud for particle size smaller than 2~$\upmu$m.
        \item We have derived vertical profiles of water ice clouds opacity and average particle size
            before and during the MY 34 GDS (from $L_s~=~165^\circ$ to
            $L_s~=~243^\circ$).
        \item We detect water ice clouds in most of the observations across the planet
            except during the onset of the GDS (4/5 of our profiles show at least one detection
            of water ice, and half of the non-detections are located within the onset of the GDS).
        \item We retrieve very high haze top altitudes during the GDS, notably at solar longitudes close to the onset with 
			evidence for mesospheric clouds at altitudes $\geq90$~km, and even up to 100~km for the maximum haze top altitude. 
			This suggests that GDS may elevates peak haze altitudes by 10 to 20~km compared to the typical perihelion storm 
			season values.
        \item We observe that there is a general trend of decreasing particle size with altitude for the whole dataset, 
			and more specifically that the particle size within a given cloud/profile also decreases with altitude, 
			typically from 1 to 2~$\upmu$m at the bottom of the cloud down to less than 0.2~$\upmu$m at the top.
        \item We note a decrease in the mean water ice particle size during the GDS, with a median $r_\mathrm{eff}$ of 
			0.3~$\upmu$m for water ice particles before the storm that become $\leq$ 0.1~$\upmu$m during the dust storm. 
			This is related to an increase of the average altitude of clouds during the GDS, 
			with a higher proportion of mesospheric, small-grained water ice clouds.
        \item Finally, we observe evidence for large water ice particles ($r_\mathrm{eff}\sim1.5-2~\upmu$m) at unexpectedly high 
            altitudes during the GDS (between 50~km and 70~km).
    \end{enumerate}

    To conclude, this study shows that the MY~34 GDS has impacted the water ice clouds distribution,
    with more frequent clouds detections at very high altitudes up to above 90~km, along with evidence of unexpected particles with radii $\geq 1.5~\upmu$m up to 70~km.

% \appendix
% \section{Supplementary materials}

\acknowledgments
ExoMars is a space mission of ESA and Roscosmos.
The ACS experiment is led by IKI Space Research Institute in Moscow.
The project acknowledges funding by Roscosmos and CNES.
Science operations of ACS are funded by Roscosmos and ESA.
Science support in IKI is funded by Federal agency of science organization (FANO).
MJW acknowledge the support of the D'Alembert Fellowship program.
Raw ACS data are available on the ESA PSA at 
\url{https://archives.esac.esa.int/psa/#!Table%20View/ACS=instrument}.
Derived opacity and particles sizes can be foud at
\url{http://dx.doi.org/10.17632/w7yff8r97s.1}.

%\url{https://archives.esac.esa.int/psa/#!Table%20View/ACS=instrument}

%\bibliography{biblio_JGR2019.bib}
\bibliography{biblio_JGR2019}

\begin{thebibliography}{}

\bibitem [\protect \citeauthoryear {%
Audouard%
\ \protect \BOthers {.}}{%
Audouard%
\ \protect \BOthers {.}}{%
{\protect \APACyear {2014}}%
}]{%
audouard_2014}
\APACinsertmetastar {%
audouard_2014}%
\begin{APACrefauthors}%
Audouard, J.%
, Poulet, F.%
, Vincendon, M.%
, Milliken, R\BPBI E.%
, Jouglet, D.%
, Bibring, J\BHBI P.%
\BDBL {}Langevin, Y.%
\end{APACrefauthors}%
\unskip\
\newblock
\APACrefYearMonthDay{2014}{}{}.
\newblock
{\BBOQ}\APACrefatitle {Water in the {{Martian}} Regolith from {{OMEGA}}/{{Mars
  Express}}} {Water in the {{Martian}} regolith from {{OMEGA}}/{{Mars
  Express}}}.{\BBCQ}
\newblock
\APACjournalVolNumPages{Journal of Geophysical Research:
  Planets}{119}{8}{1969--1989}.
\newblock
\begin{APACrefDOI} \doi{10.1002/2014JE004649} \end{APACrefDOI}
\PrintBackRefs{\CurrentBib}

\bibitem [\protect \citeauthoryear {%
Bevington%
\ \BBA {} Robinson%
}{%
Bevington%
\ \BBA {} Robinson%
}{%
{\protect \APACyear {1992}}%
}]{%
bevington_1992}
\APACinsertmetastar {%
bevington_1992}%
\begin{APACrefauthors}%
Bevington, P\BPBI R.%
\BCBT {}\ \BBA {} Robinson, D\BPBI K.%
\end{APACrefauthors}%
\unskip\
\newblock
\APACrefYear{1992}.
\newblock
\APACrefbtitle {Data Reduction and Error Analysis for the Physical Sciences -
  {{Second}} Edition} {Data reduction and error analysis for the physical
  sciences - {{Second}} edition}.
\newblock
\APACaddressPublisher{}{{McGraw-Hill Inc}}.
\PrintBackRefs{\CurrentBib}

\bibitem [\protect \citeauthoryear {%
Bibring%
\ \protect \BOthers {.}}{%
Bibring%
\ \protect \BOthers {.}}{%
{\protect \APACyear {2004}}%
}]{%
bibring_OMEGA_2004}
\APACinsertmetastar {%
bibring_OMEGA_2004}%
\begin{APACrefauthors}%
Bibring, J\BHBI P.%
, Soufflot, A.%
, Berthé, M.%
, Langevin, Y.%
, Gondet, B.%
, Drossart, P.%
\BDBL {}Forget, F.%
\end{APACrefauthors}%
\unskip\
\newblock
\APACrefYearMonthDay{2004}{{\APACmonth{08}}}{}.
\newblock
{\BBOQ}\APACrefatitle {{OMEGA}: {Observatoire} pour la {Min\'eralogie},
  l'{Eau}, les {Glaces} et l'{Activit\'e}} {{OMEGA}: {Observatoire} pour la
  {Min\'eralogie}, l'{Eau}, les {Glaces} et l'{Activit\'e}}.{\BBCQ}
\newblock
\APACjournalVolNumPages{ESA Publication Division}{1240}{}{37--49}.
\PrintBackRefs{\CurrentBib}

\bibitem [\protect \citeauthoryear {%
BIMP%
\ \protect \BOthers {.}}{%
BIMP%
\ \protect \BOthers {.}}{%
{\protect \APACyear {2008}}%
}]{%
JCGM_100}
\APACinsertmetastar {%
JCGM_100}%
\begin{APACrefauthors}%
BIMP%
, IEC%
, IFCC%
, ILAC%
, ISO%
, IUPAC%
\BDBL {}OIML%
\end{APACrefauthors}%
\unskip\
\newblock
\APACrefYearMonthDay{2008}{}{}.
\newblock
\APACrefbtitle {{JCGM} 100: {Evaluation} of {Measurement} {Data} - {Guide} to
  the {Expression} of {Uncertainty} in {Measurement}} {{JCGM} 100: {Evaluation}
  of {Measurement} {Data} - {Guide} to the {Expression} of {Uncertainty} in
  {Measurement}}\ \APACbVolEdTR{}{\BTR{}}.
\newblock
\APACaddressInstitution{}{JCGM}.
\PrintBackRefs{\CurrentBib}

\bibitem [\protect \citeauthoryear {%
Calvin%
}{%
Calvin%
}{%
{\protect \APACyear {1997}}%
}]{%
calvin_1997}
\APACinsertmetastar {%
calvin_1997}%
\begin{APACrefauthors}%
Calvin, W\BPBI M.%
\end{APACrefauthors}%
\unskip\
\newblock
\APACrefYearMonthDay{1997}{}{}.
\newblock
{\BBOQ}\APACrefatitle {Variation of the 3-{$\mu$}m Absorption Feature on
  {{Mars}}: {{Observations}} over Eastern {{Valles Marineris}} by the
  {{Mariner}} 6 Infrared Spectrometer} {Variation of the 3-{$\mu$}m absorption
  feature on {{Mars}}: {{Observations}} over eastern {{Valles Marineris}} by
  the {{Mariner}} 6 infrared spectrometer}.{\BBCQ}
\newblock
\APACjournalVolNumPages{Journal of Geophysical Research:
  Planets}{102}{E4}{9097-9107}.
\newblock
\begin{APACrefDOI} \doi{10.1029/96JE03767} \end{APACrefDOI}
\PrintBackRefs{\CurrentBib}

\bibitem [\protect \citeauthoryear {%
Chaffin%
, Deighan%
, Schneider%
\BCBL {}\ \BBA {} Stewart%
}{%
Chaffin%
\ \protect \BOthers {.}}{%
{\protect \APACyear {2017}}%
}]{%
chaffin_2017}
\APACinsertmetastar {%
chaffin_2017}%
\begin{APACrefauthors}%
Chaffin, M\BPBI S.%
, Deighan, J.%
, Schneider, N\BPBI M.%
\BCBL {}\ \BBA {} Stewart, A\BPBI I\BPBI F.%
\end{APACrefauthors}%
\unskip\
\newblock
\APACrefYearMonthDay{2017}{{\APACmonth{01}}}{}.
\newblock
{\BBOQ}\APACrefatitle {Elevated Atmospheric Escape of Atomic Hydrogen from
  {{Mars}} Induced by High-Altitude Water} {Elevated atmospheric escape of
  atomic hydrogen from {{Mars}} induced by high-altitude water}.{\BBCQ}
\newblock
\APACjournalVolNumPages{Nature Geoscience}{10}{}{174-178}.
\newblock
\begin{APACrefDOI} \doi{10.1038/ngeo2887} \end{APACrefDOI}
\PrintBackRefs{\CurrentBib}

\bibitem [\protect \citeauthoryear {%
Clancy%
\ \protect \BOthers {.}}{%
Clancy%
\ \protect \BOthers {.}}{%
{\protect \APACyear {1996}}%
}]{%
clancy_1996}
\APACinsertmetastar {%
clancy_1996}%
\begin{APACrefauthors}%
Clancy, R\BPBI T.%
, Grossman, A\BPBI W.%
, Wolff, M\BPBI J.%
, James, P\BPBI B.%
, Rudy, D\BPBI J.%
, Billawala, Y\BPBI N.%
\BDBL {}Muhleman, D\BPBI O.%
\end{APACrefauthors}%
\unskip\
\newblock
\APACrefYearMonthDay{1996}{{\APACmonth{07}}}{}.
\newblock
{\BBOQ}\APACrefatitle {Water {Vapor} {Saturation} at {Low} {Altitudes} around
  {Mars} {Aphelion}: {A} {Key} to {Mars} {Climate}?} {Water {Vapor}
  {Saturation} at {Low} {Altitudes} around {Mars} {Aphelion}: {A} {Key} to
  {Mars} {Climate}?}{\BBCQ}
\newblock
\APACjournalVolNumPages{Icarus}{122}{1}{36--62}.
\newblock
\begin{APACrefDOI} \doi{10.1006/icar.1996.0108} \end{APACrefDOI}
\PrintBackRefs{\CurrentBib}

\bibitem [\protect \citeauthoryear {%
Clancy%
\ \protect \BOthers {.}}{%
Clancy%
\ \protect \BOthers {.}}{%
{\protect \APACyear {2017}}%
}]{%
clancy_2017}
\APACinsertmetastar {%
clancy_2017}%
\begin{APACrefauthors}%
Clancy, R\BPBI T.%
, Sandor, B.%
, Wolff, M.%
, Lefèvre, F.%
, Navarro, T.%
, Smith, M.%
\BDBL {}Toigo, A.%
\end{APACrefauthors}%
\unskip\
\newblock
\APACrefYearMonthDay{2017}{{\APACmonth{01}}}{}.
\newblock
{\BBOQ}\APACrefatitle {The {Distribution} of {Mars} {Water} {Vapor} {Versus}
  {Altitude}, {Season}, and {Latitude} as {Derived} from {Global} {Comparisons}
  of {CRISM} {Retrieved} and {LMD} {GCM} {Simulated}
  {O}\_2({\textasciicircum}1∆\_g) {Dayglow} {Profiles}} {The {Distribution}
  of {Mars} {Water} {Vapor} {Versus} {Altitude}, {Season}, and {Latitude} as
  {Derived} from {Global} {Comparisons} of {CRISM} {Retrieved} and {LMD} {GCM}
  {Simulated} {O}\_2({\textasciicircum}1∆\_g) {Dayglow} {Profiles}}.{\BBCQ}
\newblock
\BIn{} \APACrefbtitle {The Mars Atmosphere: Modelling and observation} {The
  mars atmosphere: Modelling and observation}\ (\BPG~3203).
\PrintBackRefs{\CurrentBib}

\bibitem [\protect \citeauthoryear {%
Clancy%
\ \protect \BOthers {.}}{%
Clancy%
\ \protect \BOthers {.}}{%
{\protect \APACyear {2000}}%
}]{%
clancy_2000}
\APACinsertmetastar {%
clancy_2000}%
\begin{APACrefauthors}%
Clancy, R\BPBI T.%
, Sandor, B\BPBI J.%
, Wolff, M\BPBI J.%
, Christensen, P\BPBI R.%
, Smith, M\BPBI D.%
, Pearl, J\BPBI C.%
\BDBL {}Wilson, R\BPBI J.%
\end{APACrefauthors}%
\unskip\
\newblock
\APACrefYearMonthDay{2000}{}{}.
\newblock
{\BBOQ}\APACrefatitle {An Intercomparison of Ground-Based Millimeter, {{MGS
  TES}}, and {{Viking}} Atmospheric Temperature Measurements: {{Seasonal}} and
  Interannual Variability of Temperatures and Dust Loading in the Global
  {{Mars}} Atmosphere} {An intercomparison of ground-based millimeter, {{MGS
  TES}}, and {{Viking}} atmospheric temperature measurements: {{Seasonal}} and
  interannual variability of temperatures and dust loading in the global
  {{Mars}} atmosphere}.{\BBCQ}
\newblock
\APACjournalVolNumPages{Journal of Geophysical Research:
  Planets}{105}{E4}{9553-9571}.
\newblock
\begin{APACrefDOI} \doi{10.1029/1999JE001089} \end{APACrefDOI}
\PrintBackRefs{\CurrentBib}

\bibitem [\protect \citeauthoryear {%
Clancy%
\ \protect \BOthers {.}}{%
Clancy%
\ \protect \BOthers {.}}{%
{\protect \APACyear {2019}}%
}]{%
clancy_2019}
\APACinsertmetastar {%
clancy_2019}%
\begin{APACrefauthors}%
Clancy, R\BPBI T.%
, Wolff, M\BPBI J.%
, Smith, M\BPBI D.%
, Kleinb\"ohl, A.%
, Cantor, B\BPBI A.%
, Murchie, S\BPBI L.%
\BDBL {}Sandor, B\BPBI J.%
\end{APACrefauthors}%
\unskip\
\newblock
\APACrefYearMonthDay{2019}{{\APACmonth{08}}}{}.
\newblock
{\BBOQ}\APACrefatitle {The distribution, composition, and particle properties
  of {Mars} mesospheric aerosols: {An} analysis of {CRISM} visible/near-{IR}
  limb spectra with context from near-coincident {MCS} and {MARCI}
  observations} {The distribution, composition, and particle properties of
  {Mars} mesospheric aerosols: {An} analysis of {CRISM} visible/near-{IR} limb
  spectra with context from near-coincident {MCS} and {MARCI}
  observations}.{\BBCQ}
\newblock
\APACjournalVolNumPages{Icarus}{328}{}{246--273}.
\newblock
\begin{APACrefDOI} \doi{10.1016/j.icarus.2019.03.025} \end{APACrefDOI}
\PrintBackRefs{\CurrentBib}

\bibitem [\protect \citeauthoryear {%
Clancy%
\ \protect \BOthers {.}}{%
Clancy%
\ \protect \BOthers {.}}{%
{\protect \APACyear {2010}}%
}]{%
clancy_2010}
\APACinsertmetastar {%
clancy_2010}%
\begin{APACrefauthors}%
Clancy, R\BPBI T.%
, Wolff, M\BPBI J.%
, Whitney, B\BPBI A.%
, Cantor, B\BPBI A.%
, Smith, M\BPBI D.%
\BCBL {}\ \BBA {} McConnochie, T\BPBI H.%
\end{APACrefauthors}%
\unskip\
\newblock
\APACrefYearMonthDay{2010}{{\APACmonth{05}}}{}.
\newblock
{\BBOQ}\APACrefatitle {Extension of Atmospheric Dust Loading to High Altitudes
  during the 2001 {{Mars}} Dust Storm: {{MGS TES}} Limb Observations}
  {Extension of atmospheric dust loading to high altitudes during the 2001
  {{Mars}} dust storm: {{MGS TES}} limb observations}.{\BBCQ}
\newblock
\APACjournalVolNumPages{Icarus}{207}{1}{98-109}.
\newblock
\begin{APACrefDOI} \doi{10.1016/j.icarus.2009.10.011} \end{APACrefDOI}
\PrintBackRefs{\CurrentBib}

\bibitem [\protect \citeauthoryear {%
Crismani%
\ \protect \BOthers {.}}{%
Crismani%
\ \protect \BOthers {.}}{%
{\protect \APACyear {2017}}%
}]{%
crismani_2017}
\APACinsertmetastar {%
crismani_2017}%
\begin{APACrefauthors}%
Crismani, M\BPBI M\BPBI J.%
, Schneider, N\BPBI M.%
, Plane, J\BPBI M\BPBI C.%
, Evans, J\BPBI S.%
, Jain, S\BPBI K.%
, Chaffin, M\BPBI S.%
\BDBL {}Jakosky, B\BPBI M.%
\end{APACrefauthors}%
\unskip\
\newblock
\APACrefYearMonthDay{2017}{{\APACmonth{06}}}{}.
\newblock
{\BBOQ}\APACrefatitle {Detection of a Persistent Meteoric Metal Layer in the
  {{Martian}} Atmosphere} {Detection of a persistent meteoric metal layer in
  the {{Martian}} atmosphere}.{\BBCQ}
\newblock
\APACjournalVolNumPages{Nature Geoscience}{10}{}{401-404}.
\newblock
\begin{APACrefDOI} \doi{10.1038/ngeo2958} \end{APACrefDOI}
\PrintBackRefs{\CurrentBib}

\bibitem [\protect \citeauthoryear {%
Fedorova%
\ \protect \BOthers {.}}{%
Fedorova%
\ \protect \BOthers {.}}{%
{\protect \APACyear {2018}}%
}]{%
fedorova_2018}
\APACinsertmetastar {%
fedorova_2018}%
\begin{APACrefauthors}%
Fedorova, A.%
, Bertaux, J\BHBI L.%
, Betsis, D.%
, Montmessin, F.%
, Korablev, O.%
, Maltagliati, L.%
\BCBL {}\ \BBA {} Clarke, J.%
\end{APACrefauthors}%
\unskip\
\newblock
\APACrefYearMonthDay{2018}{{\APACmonth{01}}}{}.
\newblock
{\BBOQ}\APACrefatitle {Water vapor in the middle atmosphere of {Mars} during
  the 2007 global dust storm} {Water vapor in the middle atmosphere of {Mars}
  during the 2007 global dust storm}.{\BBCQ}
\newblock
\APACjournalVolNumPages{Icarus}{300}{}{440--457}.
\newblock
\begin{APACrefDOI} \doi{10.1016/j.icarus.2017.09.025} \end{APACrefDOI}
\PrintBackRefs{\CurrentBib}

\bibitem [\protect \citeauthoryear {%
Fedorova%
\ \protect \BOthers {.}}{%
Fedorova%
\ \protect \BOthers {.}}{%
{\protect \APACyear {2020}}%
}]{%
fedorova_2020}
\APACinsertmetastar {%
fedorova_2020}%
\begin{APACrefauthors}%
Fedorova, A.%
, Montmessin, F.%
, Korablev, O.%
, Luginin, M.%
, Trokhimovskiy, A.%
, Belyaev, D\BPBI A.%
\BDBL {}Wilson, C\BPBI F.%
\end{APACrefauthors}%
\unskip\
\newblock
\APACrefYearMonthDay{2020}{{\APACmonth{01}}}{}.
\newblock
{\BBOQ}\APACrefatitle {Stormy Water on {{Mars}}: {{The}} Distribution and
  Saturation of Atmospheric Water during the Dusty Season} {Stormy water on
  {{Mars}}: {{The}} distribution and saturation of atmospheric water during the
  dusty season}.{\BBCQ}
\newblock
\APACjournalVolNumPages{Science}{}{}{eaay9522}.
\newblock
\begin{APACrefDOI} \doi{10.1126/science.aay9522} \end{APACrefDOI}
\PrintBackRefs{\CurrentBib}

\bibitem [\protect \citeauthoryear {%
Forget%
\ \protect \BOthers {.}}{%
Forget%
\ \protect \BOthers {.}}{%
{\protect \APACyear {1999}}%
}]{%
forget_1999}
\APACinsertmetastar {%
forget_1999}%
\begin{APACrefauthors}%
Forget, F.%
, Hourdin, F.%
, Fournier, R.%
, Hourdin, C.%
, Talagrand, O.%
, Collins, M.%
\BDBL {}Huot, J\BHBI P.%
\end{APACrefauthors}%
\unskip\
\newblock
\APACrefYearMonthDay{1999}{}{}.
\newblock
{\BBOQ}\APACrefatitle {Improved General Circulation Models of the {{Martian}}
  Atmosphere from the Surface to above 80 Km} {Improved general circulation
  models of the {{Martian}} atmosphere from the surface to above 80 km}.{\BBCQ}
\newblock
\APACjournalVolNumPages{Journal of Geophysical Research:
  Planets}{104}{E10}{24155-24175}.
\newblock
\begin{APACrefDOI} \doi{10.1029/1999JE001025} \end{APACrefDOI}
\PrintBackRefs{\CurrentBib}

\bibitem [\protect \citeauthoryear {%
Goldman%
\ \BBA {} Saunders%
}{%
Goldman%
\ \BBA {} Saunders%
}{%
{\protect \APACyear {1979}}%
}]{%
goldman_saunders_1979}
\APACinsertmetastar {%
goldman_saunders_1979}%
\begin{APACrefauthors}%
Goldman, A.%
\BCBT {}\ \BBA {} Saunders, R.%
\end{APACrefauthors}%
\unskip\
\newblock
\APACrefYearMonthDay{1979}{{\APACmonth{02}}}{}.
\newblock
{\BBOQ}\APACrefatitle {Analysis of atmospheric infrared spectra for altitude
  distribution of atmospheric trace constituents—{I}. {Method} of analysis}
  {Analysis of atmospheric infrared spectra for altitude distribution of
  atmospheric trace constituents—{I}. {Method} of analysis}.{\BBCQ}
\newblock
\APACjournalVolNumPages{Journal of Quantitative Spectroscopy and Radiative
  Transfer}{21}{2}{155--161}.
\newblock
\begin{APACrefDOI} \doi{10.1016/0022-4073(79)90027-X} \end{APACrefDOI}
\PrintBackRefs{\CurrentBib}

\bibitem [\protect \citeauthoryear {%
Gooding%
}{%
Gooding%
}{%
{\protect \APACyear {1986}}%
}]{%
gooding_1986}
\APACinsertmetastar {%
gooding_1986}%
\begin{APACrefauthors}%
Gooding, J\BPBI L.%
\end{APACrefauthors}%
\unskip\
\newblock
\APACrefYearMonthDay{1986}{{\APACmonth{04}}}{}.
\newblock
{\BBOQ}\APACrefatitle {Martian Dust Particles as Condensation Nuclei: {{A}}
  Preliminary Assessment of Mineralogical Factors} {Martian dust particles as
  condensation nuclei: {{A}} preliminary assessment of mineralogical
  factors}.{\BBCQ}
\newblock
\APACjournalVolNumPages{Icarus}{66}{1}{56-74}.
\newblock
\begin{APACrefDOI} \doi{10.1016/0019-1035(86)90006-0} \end{APACrefDOI}
\PrintBackRefs{\CurrentBib}

\bibitem [\protect \citeauthoryear {%
Guzewich%
\ \protect \BOthers {.}}{%
Guzewich%
\ \protect \BOthers {.}}{%
{\protect \APACyear {2019}}%
}]{%
guzewich_GDS_2019}
\APACinsertmetastar {%
guzewich_GDS_2019}%
\begin{APACrefauthors}%
Guzewich, S\BPBI D.%
, Lemmon, M.%
, Smith, C\BPBI L.%
, Mart\'inez, G.%
, Vicente‐Retortillo, Ã\BPBI d.%
, Newman, C\BPBI E.%
\BDBL {}Mier, M\BHBI P\BPBI Z.%
\end{APACrefauthors}%
\unskip\
\newblock
\APACrefYearMonthDay{2019}{{\APACmonth{01}}}{}.
\newblock
{\BBOQ}\APACrefatitle {Mars {Science} {Laboratory} {Observations} of the
  2018/{Mars} {Year} 34 {Global} {Dust} {Storm}} {Mars {Science} {Laboratory}
  {Observations} of the 2018/{Mars} {Year} 34 {Global} {Dust} {Storm}}.{\BBCQ}
\newblock
\APACjournalVolNumPages{Geophysical Research Letters}{46}{}{71--79}.
\newblock
\begin{APACrefDOI} \doi{10.1029/2018GL080839} \end{APACrefDOI}
\PrintBackRefs{\CurrentBib}

\bibitem [\protect \citeauthoryear {%
Guzewich%
\ \BBA {} Smith%
}{%
Guzewich%
\ \BBA {} Smith%
}{%
{\protect \APACyear {2019}}%
}]{%
guzewich_2019}
\APACinsertmetastar {%
guzewich_2019}%
\begin{APACrefauthors}%
Guzewich, S\BPBI D.%
\BCBT {}\ \BBA {} Smith, M\BPBI D.%
\end{APACrefauthors}%
\unskip\
\newblock
\APACrefYearMonthDay{2019}{{\APACmonth{02}}}{}.
\newblock
{\BBOQ}\APACrefatitle {Seasonal {Variation} in {Martian} {Water} {Ice} {Cloud}
  {Particle} {Size}} {Seasonal {Variation} in {Martian} {Water} {Ice} {Cloud}
  {Particle} {Size}}.{\BBCQ}
\newblock
\APACjournalVolNumPages{Journal of Geophysical Research:
  Planets}{124}{2}{636--643}.
\newblock
\begin{APACrefDOI} \doi{10.1029/2018JE005843} \end{APACrefDOI}
\PrintBackRefs{\CurrentBib}

\bibitem [\protect \citeauthoryear {%
Guzewich%
, Smith%
\BCBL {}\ \BBA {} Wolff%
}{%
Guzewich%
\ \protect \BOthers {.}}{%
{\protect \APACyear {2014}}%
}]{%
guzewich_2014}
\APACinsertmetastar {%
guzewich_2014}%
\begin{APACrefauthors}%
Guzewich, S\BPBI D.%
, Smith, M\BPBI D.%
\BCBL {}\ \BBA {} Wolff, M\BPBI J.%
\end{APACrefauthors}%
\unskip\
\newblock
\APACrefYearMonthDay{2014}{{\APACmonth{12}}}{}.
\newblock
{\BBOQ}\APACrefatitle {The vertical distribution of {Martian} aerosol particle
  size: {Vertical} {Profile} of {Mars} {Aerosol} {Size}} {The vertical
  distribution of {Martian} aerosol particle size: {Vertical} {Profile} of
  {Mars} {Aerosol} {Size}}.{\BBCQ}
\newblock
\APACjournalVolNumPages{Journal of Geophysical Research:
  Planets}{119}{12}{2694--2708}.
\newblock
\begin{APACrefDOI} \doi{10.1002/2014JE004704} \end{APACrefDOI}
\PrintBackRefs{\CurrentBib}

\bibitem [\protect \citeauthoryear {%
Hanel%
\ \protect \BOthers {.}}{%
Hanel%
\ \protect \BOthers {.}}{%
{\protect \APACyear {1972}}%
}]{%
hanel_1972}
\APACinsertmetastar {%
hanel_1972}%
\begin{APACrefauthors}%
Hanel, R.%
, Conrath, B.%
, Hovis, W.%
, Kunde, V.%
, Lowman, P.%
, Maguire, W.%
\BDBL {}Burke, T.%
\end{APACrefauthors}%
\unskip\
\newblock
\APACrefYearMonthDay{1972}{{\APACmonth{10}}}{}.
\newblock
{\BBOQ}\APACrefatitle {Investigation of the {Martian} environment by infrared
  spectroscopy on {Mariner} 9} {Investigation of the {Martian} environment by
  infrared spectroscopy on {Mariner} 9}.{\BBCQ}
\newblock
\APACjournalVolNumPages{Icarus}{17}{2}{423--442}.
\newblock
\begin{APACrefDOI} \doi{10.1016/0019-1035(72)90009-7} \end{APACrefDOI}
\PrintBackRefs{\CurrentBib}

\bibitem [\protect \citeauthoryear {%
Hansen%
\ \BBA {} Travis%
}{%
Hansen%
\ \BBA {} Travis%
}{%
{\protect \APACyear {1974}}%
}]{%
hansen_1974}
\APACinsertmetastar {%
hansen_1974}%
\begin{APACrefauthors}%
Hansen, J\BPBI E.%
\BCBT {}\ \BBA {} Travis, L\BPBI D.%
\end{APACrefauthors}%
\unskip\
\newblock
\APACrefYearMonthDay{1974}{{\APACmonth{10}}}{}.
\newblock
{\BBOQ}\APACrefatitle {Light scattering in planetary atmospheres} {Light
  scattering in planetary atmospheres}.{\BBCQ}
\newblock
\APACjournalVolNumPages{Space Science Reviews}{16}{4}{527--610}.
\newblock
\begin{APACrefDOI} \doi{10.1007/BF00168069} \end{APACrefDOI}
\PrintBackRefs{\CurrentBib}

\bibitem [\protect \citeauthoryear {%
Hartwick%
, Toon%
\BCBL {}\ \BBA {} Heavens%
}{%
Hartwick%
\ \protect \BOthers {.}}{%
{\protect \APACyear {2019}}%
}]{%
hartwick_2019}
\APACinsertmetastar {%
hartwick_2019}%
\begin{APACrefauthors}%
Hartwick, V\BPBI L.%
, Toon, O\BPBI B.%
\BCBL {}\ \BBA {} Heavens, N\BPBI G.%
\end{APACrefauthors}%
\unskip\
\newblock
\APACrefYearMonthDay{2019}{{\APACmonth{07}}}{}.
\newblock
{\BBOQ}\APACrefatitle {High-altitude water ice cloud formation on {Mars}
  controlled by interplanetary dust particles} {High-altitude water ice cloud
  formation on {Mars} controlled by interplanetary dust particles}.{\BBCQ}
\newblock
\APACjournalVolNumPages{Nature Geoscience}{12}{7}{516--521}.
\newblock
\begin{APACrefDOI} \doi{10.1038/s41561-019-0379-6} \end{APACrefDOI}
\PrintBackRefs{\CurrentBib}

\bibitem [\protect \citeauthoryear {%
Heavens%
\ \protect \BOthers {.}}{%
Heavens%
\ \protect \BOthers {.}}{%
{\protect \APACyear {2018}}%
}]{%
heavens_2018}
\APACinsertmetastar {%
heavens_2018}%
\begin{APACrefauthors}%
Heavens, N\BPBI G.%
, Kleinböhl, A.%
, Chaffin, M\BPBI S.%
, Halekas, J\BPBI S.%
, Kass, D\BPBI M.%
, Hayne, P\BPBI O.%
\BDBL {}Schofield, J\BPBI T.%
\end{APACrefauthors}%
\unskip\
\newblock
\APACrefYearMonthDay{2018}{{\APACmonth{02}}}{}.
\newblock
{\BBOQ}\APACrefatitle {Hydrogen escape from {Mars} enhanced by deep convection
  in dust storms} {Hydrogen escape from {Mars} enhanced by deep convection in
  dust storms}.{\BBCQ}
\newblock
\APACjournalVolNumPages{Nature Astronomy}{2}{2}{126--132}.
\newblock
\begin{APACrefDOI} \doi{10.1038/s41550-017-0353-4} \end{APACrefDOI}
\PrintBackRefs{\CurrentBib}

\bibitem [\protect \citeauthoryear {%
Heavens%
\ \protect \BOthers {.}}{%
Heavens%
\ \protect \BOthers {.}}{%
{\protect \APACyear {2011}}%
}]{%
heavens_2011a}
\APACinsertmetastar {%
heavens_2011a}%
\begin{APACrefauthors}%
Heavens, N\BPBI G.%
, Richardson, M\BPBI I.%
, Kleinb{\"o}hl, A.%
, Kass, D\BPBI M.%
, McCleese, D\BPBI J.%
, Abdou, W.%
\BDBL {}Wolkenberg, P\BPBI M.%
\end{APACrefauthors}%
\unskip\
\newblock
\APACrefYearMonthDay{2011}{}{}.
\newblock
{\BBOQ}\APACrefatitle {The Vertical Distribution of Dust in the {{Martian}}
  Atmosphere during Northern Spring and Summer: {{Observations}} by the {{Mars
  Climate Sounder}} and Analysis of Zonal Average Vertical Dust Profiles} {The
  vertical distribution of dust in the {{Martian}} atmosphere during northern
  spring and summer: {{Observations}} by the {{Mars Climate Sounder}} and
  analysis of zonal average vertical dust profiles}.{\BBCQ}
\newblock
\APACjournalVolNumPages{Journal of Geophysical Research: Planets}{116}{E4}{}.
\newblock
\begin{APACrefDOI} \doi{10.1029/2010JE003691} \end{APACrefDOI}
\PrintBackRefs{\CurrentBib}

\bibitem [\protect \citeauthoryear {%
Jaquin%
, Gierasch%
\BCBL {}\ \BBA {} Kahn%
}{%
Jaquin%
\ \protect \BOthers {.}}{%
{\protect \APACyear {1986}}%
}]{%
jaquin_1986}
\APACinsertmetastar {%
jaquin_1986}%
\begin{APACrefauthors}%
Jaquin, F.%
, Gierasch, P.%
\BCBL {}\ \BBA {} Kahn, R.%
\end{APACrefauthors}%
\unskip\
\newblock
\APACrefYearMonthDay{1986}{{\APACmonth{12}}}{}.
\newblock
{\BBOQ}\APACrefatitle {The Vertical Structure of Limb Hazes in the {{Martian}}
  Atmosphere} {The vertical structure of limb hazes in the {{Martian}}
  atmosphere}.{\BBCQ}
\newblock
\APACjournalVolNumPages{Icarus}{68}{3}{442-461}.
\newblock
\begin{APACrefDOI} \doi{10.1016/0019-1035(86)90050-3} \end{APACrefDOI}
\PrintBackRefs{\CurrentBib}

\bibitem [\protect \citeauthoryear {%
Jouglet%
\ \protect \BOthers {.}}{%
Jouglet%
\ \protect \BOthers {.}}{%
{\protect \APACyear {2007}}%
}]{%
jouglet_2007}
\APACinsertmetastar {%
jouglet_2007}%
\begin{APACrefauthors}%
Jouglet, D.%
, Poulet, F.%
, Milliken, R\BPBI E.%
, Mustard, J\BPBI F.%
, Bibring, J\BHBI P.%
, Langevin, Y.%
\BDBL {}Gomez, C.%
\end{APACrefauthors}%
\unskip\
\newblock
\APACrefYearMonthDay{2007}{}{}.
\newblock
{\BBOQ}\APACrefatitle {Hydration state of the {Martian} surface as seen by
  {Mars} {Express} {OMEGA}: 1. {Analysis} of the 3 μm hydration feature}
  {Hydration state of the {Martian} surface as seen by {Mars} {Express}
  {OMEGA}: 1. {Analysis} of the 3 μm hydration feature}.{\BBCQ}
\newblock
\APACjournalVolNumPages{Journal of Geophysical Research: Planets}{112}{E8}{}.
\newblock
\begin{APACrefDOI} \doi{10.1029/2006JE002846} \end{APACrefDOI}
\PrintBackRefs{\CurrentBib}

\bibitem [\protect \citeauthoryear {%
D.~Kass%
\ \protect \BOthers {.}}{%
D.~Kass%
\ \protect \BOthers {.}}{%
{\protect \APACyear {2019}}%
}]{%
kass_2019}
\APACinsertmetastar {%
kass_2019}%
\begin{APACrefauthors}%
Kass, D.%
, Schofield, J.%
, Kleinb{\"o}hl, A.%
, McCleese, D.%
, Heavens, N.%
, Shirley, J.%
\BCBL {}\ \BBA {} Steele, L.%
\end{APACrefauthors}%
\unskip\
\newblock
\APACrefYearMonthDay{2019}{{\APACmonth{09}}}{}.
\newblock
{\BBOQ}\APACrefatitle {Mars {{Climate Sounder}} Observation of {{Mars}}' 2018
  Global Dust Storm} {Mars {{Climate Sounder}} observation of {{Mars}}' 2018
  global dust storm}.{\BBCQ}
\newblock
\APACjournalVolNumPages{Geophysical Research Letters}{}{}{}.
\newblock
\begin{APACrefDOI} \doi{10.1029/2019GL083931} \end{APACrefDOI}
\PrintBackRefs{\CurrentBib}

\bibitem [\protect \citeauthoryear {%
D\BPBI M.~Kass%
, Kleinb{\"o}hl%
, McCleese%
, Schofield%
\BCBL {}\ \BBA {} Smith%
}{%
D\BPBI M.~Kass%
\ \protect \BOthers {.}}{%
{\protect \APACyear {2016}}%
}]{%
kass_2016}
\APACinsertmetastar {%
kass_2016}%
\begin{APACrefauthors}%
Kass, D\BPBI M.%
, Kleinb{\"o}hl, A.%
, McCleese, D\BPBI J.%
, Schofield, J\BPBI T.%
\BCBL {}\ \BBA {} Smith, M\BPBI D.%
\end{APACrefauthors}%
\unskip\
\newblock
\APACrefYearMonthDay{2016}{{\APACmonth{06}}}{}.
\newblock
{\BBOQ}\APACrefatitle {Interannual Similarity in the {{Martian}} Atmosphere
  during the Dust Storm Season} {Interannual similarity in the {{Martian}}
  atmosphere during the dust storm season}.{\BBCQ}
\newblock
\APACjournalVolNumPages{Geophysical Research Letters}{43}{12}{6111--6118}.
\newblock
\begin{APACrefDOI} \doi{10.1002/2016GL068978} \end{APACrefDOI}
\PrintBackRefs{\CurrentBib}

\bibitem [\protect \citeauthoryear {%
Kleinb\"ohl%
\ \protect \BOthers {.}}{%
Kleinb\"ohl%
\ \protect \BOthers {.}}{%
{\protect \APACyear {2009}}%
}]{%
kleinbohl_2009}
\APACinsertmetastar {%
kleinbohl_2009}%
\begin{APACrefauthors}%
Kleinb\"ohl, A.%
, Schofield, J\BPBI T.%
, Kass, D\BPBI M.%
, Abdou, W\BPBI A.%
, Backus, C\BPBI R.%
, Sen, B.%
\BDBL {}McCleese, D\BPBI J.%
\end{APACrefauthors}%
\unskip\
\newblock
\APACrefYearMonthDay{2009}{{\APACmonth{10}}}{}.
\newblock
{\BBOQ}\APACrefatitle {Mars {Climate} {Sounder} limb profile retrieval of
  atmospheric temperature, pressure, and dust and water ice opacity} {Mars
  {Climate} {Sounder} limb profile retrieval of atmospheric temperature,
  pressure, and dust and water ice opacity}.{\BBCQ}
\newblock
\APACjournalVolNumPages{Journal of Geophysical Research: Planets}{114}{E10}{}.
\newblock
\begin{APACrefDOI} \doi{10.1029/2009JE003358} \end{APACrefDOI}
\PrintBackRefs{\CurrentBib}

\bibitem [\protect \citeauthoryear {%
Korablev%
\ \protect \BOthers {.}}{%
Korablev%
\ \protect \BOthers {.}}{%
{\protect \APACyear {2018}}%
}]{%
korablev_ACS}
\APACinsertmetastar {%
korablev_ACS}%
\begin{APACrefauthors}%
Korablev, O.%
, Montmessin, F.%
, Trokhimovskiy, A.%
, Fedorova, A\BPBI A.%
, Shakun, A\BPBI V.%
, Grigoriev, A\BPBI V.%
\BDBL {}Zorzano, M\BPBI P.%
\end{APACrefauthors}%
\unskip\
\newblock
\APACrefYearMonthDay{2018}{{\APACmonth{02}}}{}.
\newblock
{\BBOQ}\APACrefatitle {The {Atmospheric} {Chemistry} {Suite} ({ACS}) of {Three}
  {Spectrometers} for the {ExoMars} 2016 {Trace} {Gas} {Orbiter}} {The
  {Atmospheric} {Chemistry} {Suite} ({ACS}) of {Three} {Spectrometers} for the
  {ExoMars} 2016 {Trace} {Gas} {Orbiter}}.{\BBCQ}
\newblock
\APACjournalVolNumPages{Space Science Reviews}{214}{}{7}.
\newblock
\begin{APACrefDOI} \doi{10.1007/s11214-017-0437-6} \end{APACrefDOI}
\PrintBackRefs{\CurrentBib}

\bibitem [\protect \citeauthoryear {%
Korablev%
\ \protect \BOthers {.}}{%
Korablev%
\ \protect \BOthers {.}}{%
{\protect \APACyear {2019}}%
}]{%
nature_methane}
\APACinsertmetastar {%
nature_methane}%
\begin{APACrefauthors}%
Korablev, O.%
, Vandaele, A\BPBI C.%
, Montmessin, F.%
, Fedorova, A\BPBI A.%
, Trokhimovskiy, A.%
, Forget, F\BPBI c.%
\BDBL {}{The ACS and NOMAD Science Teams}%
\end{APACrefauthors}%
\unskip\
\newblock
\APACrefYearMonthDay{2019}{{\APACmonth{04}}}{}.
\newblock
{\BBOQ}\APACrefatitle {No detection of methane on {Mars} from early {ExoMars}
  {Trace} {Gas} {Orbiter} observations} {No detection of methane on {Mars} from
  early {ExoMars} {Trace} {Gas} {Orbiter} observations}.{\BBCQ}
\newblock
\APACjournalVolNumPages{Nature}{568}{}{517--520}.
\newblock
\begin{APACrefDOI} \doi{10.1038/s41586-019-1096-4} \end{APACrefDOI}
\PrintBackRefs{\CurrentBib}

\bibitem [\protect \citeauthoryear {%
Lemmon%
\ \protect \BOthers {.}}{%
Lemmon%
\ \protect \BOthers {.}}{%
{\protect \APACyear {2019}}%
}]{%
lemmon_2019}
\APACinsertmetastar {%
lemmon_2019}%
\begin{APACrefauthors}%
Lemmon, M\BPBI T.%
, Guzewich, S\BPBI D.%
, McConnochie, T.%
, Vicente-Retortillo, A.%
, Mart{\'i}nez, G.%
, Smith, M\BPBI D.%
\BDBL {}Jacob, S.%
\end{APACrefauthors}%
\unskip\
\newblock
\APACrefYearMonthDay{2019}{{\APACmonth{08}}}{}.
\newblock
{\BBOQ}\APACrefatitle {Large {{Dust Aerosol Sizes Seen During}} the 2018
  {{Martian Global Dust Event}} by the {{Curiosity Rover}}} {Large {{Dust
  Aerosol Sizes Seen During}} the 2018 {{Martian Global Dust Event}} by the
  {{Curiosity Rover}}}.{\BBCQ}
\newblock
\APACjournalVolNumPages{Geophysical Research Letters}{}{}{}.
\newblock
\begin{APACrefDOI} \doi{10.1029/2019GL084407} \end{APACrefDOI}
\PrintBackRefs{\CurrentBib}

\bibitem [\protect \citeauthoryear {%
Madeleine%
\ \protect \BOthers {.}}{%
Madeleine%
\ \protect \BOthers {.}}{%
{\protect \APACyear {2012}}%
}]{%
madeleine_2012}
\APACinsertmetastar {%
madeleine_2012}%
\begin{APACrefauthors}%
Madeleine, J\BHBI B.%
, Forget, F.%
, Spiga, A.%
, Wolff, M\BPBI J.%
, Montmessin, F.%
, Vincendon, M.%
\BDBL {}Schmitt, B.%
\end{APACrefauthors}%
\unskip\
\newblock
\APACrefYearMonthDay{2012}{{\APACmonth{11}}}{}.
\newblock
{\BBOQ}\APACrefatitle {Aphelion water-ice cloud mapping and property retrieval
  using the {OMEGA} imaging spectrometer onboard {Mars} {Express}: {OMEGA}
  {ANALYSIS} {OF} {MARS} {WATER} {ICE} {CLOUDS}} {Aphelion water-ice cloud
  mapping and property retrieval using the {OMEGA} imaging spectrometer onboard
  {Mars} {Express}: {OMEGA} {ANALYSIS} {OF} {MARS} {WATER} {ICE}
  {CLOUDS}}.{\BBCQ}
\newblock
\APACjournalVolNumPages{Journal of Geophysical Research: Planets}{117}{E11}{}.
\newblock
\begin{APACrefDOI} \doi{10.1029/2011JE003940} \end{APACrefDOI}
\PrintBackRefs{\CurrentBib}

\bibitem [\protect \citeauthoryear {%
Maltagliati%
\ \protect \BOthers {.}}{%
Maltagliati%
\ \protect \BOthers {.}}{%
{\protect \APACyear {2011}}%
}]{%
maltagliati_2011}
\APACinsertmetastar {%
maltagliati_2011}%
\begin{APACrefauthors}%
Maltagliati, L.%
, Montmessin, F.%
, Fedorova, A.%
, Korablev, O.%
, Forget, F.%
\BCBL {}\ \BBA {} Bertaux, J\BHBI L.%
\end{APACrefauthors}%
\unskip\
\newblock
\APACrefYearMonthDay{2011}{{\APACmonth{09}}}{}.
\newblock
{\BBOQ}\APACrefatitle {Evidence of {Water} {Vapor} in {Excess} of {Saturation}
  in the {Atmosphere} of {Mars}} {Evidence of {Water} {Vapor} in {Excess} of
  {Saturation} in the {Atmosphere} of {Mars}}.{\BBCQ}
\newblock
\APACjournalVolNumPages{Science}{333}{6051}{1868--1871}.
\newblock
\begin{APACrefDOI} \doi{10.1126/science.1207957} \end{APACrefDOI}
\PrintBackRefs{\CurrentBib}

\bibitem [\protect \citeauthoryear {%
Michelangeli%
, Toon%
, Haberle%
\BCBL {}\ \BBA {} Pollack%
}{%
Michelangeli%
\ \protect \BOthers {.}}{%
{\protect \APACyear {1993}}%
}]{%
michelangeli_1993}
\APACinsertmetastar {%
michelangeli_1993}%
\begin{APACrefauthors}%
Michelangeli, D\BPBI V.%
, Toon, O\BPBI B.%
, Haberle, R\BPBI M.%
\BCBL {}\ \BBA {} Pollack, J\BPBI B.%
\end{APACrefauthors}%
\unskip\
\newblock
\APACrefYearMonthDay{1993}{{\APACmonth{04}}}{}.
\newblock
{\BBOQ}\APACrefatitle {Numerical {{Simulations}} of the {{Formation}} and
  {{Evolution}} of {{Water Ice Clouds}} in the {{Martian Atmosphere}}}
  {Numerical {{Simulations}} of the {{Formation}} and {{Evolution}} of {{Water
  Ice Clouds}} in the {{Martian Atmosphere}}}.{\BBCQ}
\newblock
\APACjournalVolNumPages{Icarus}{102}{2}{261-285}.
\newblock
\begin{APACrefDOI} \doi{10.1006/icar.1993.1048} \end{APACrefDOI}
\PrintBackRefs{\CurrentBib}

\bibitem [\protect \citeauthoryear {%
Montmessin%
, Forget%
, Rannou%
, Cabane%
\BCBL {}\ \BBA {} Haberle%
}{%
Montmessin%
\ \protect \BOthers {.}}{%
{\protect \APACyear {2004}}%
}]{%
montmessin_2004}
\APACinsertmetastar {%
montmessin_2004}%
\begin{APACrefauthors}%
Montmessin, F.%
, Forget, F.%
, Rannou, P.%
, Cabane, M.%
\BCBL {}\ \BBA {} Haberle, R\BPBI M.%
\end{APACrefauthors}%
\unskip\
\newblock
\APACrefYearMonthDay{2004}{}{}.
\newblock
{\BBOQ}\APACrefatitle {Origin and role of water ice clouds in the {Martian}
  water cycle as inferred from a general circulation model} {Origin and role of
  water ice clouds in the {Martian} water cycle as inferred from a general
  circulation model}.{\BBCQ}
\newblock
\APACjournalVolNumPages{Journal of Geophysical Research: Planets}{109}{E10}{}.
\newblock
\begin{APACrefDOI} \doi{10.1029/2004JE002284} \end{APACrefDOI}
\PrintBackRefs{\CurrentBib}

\bibitem [\protect \citeauthoryear {%
Montmessin%
\ \protect \BOthers {.}}{%
Montmessin%
\ \protect \BOthers {.}}{%
{\protect \APACyear {2006}}%
}]{%
montmessin_2006}
\APACinsertmetastar {%
montmessin_2006}%
\begin{APACrefauthors}%
Montmessin, F.%
, Qu{\'e}merais, E.%
, Bertaux, J\BPBI L.%
, Korablev, O.%
, Rannou, P.%
\BCBL {}\ \BBA {} Lebonnois, S.%
\end{APACrefauthors}%
\unskip\
\newblock
\APACrefYearMonthDay{2006}{}{}.
\newblock
{\BBOQ}\APACrefatitle {Stellar Occultations at {{UV}} Wavelengths by the
  {{SPICAM}} Instrument: {{Retrieval}} and Analysis of {{Martian}} Haze
  Profiles} {Stellar occultations at {{UV}} wavelengths by the {{SPICAM}}
  instrument: {{Retrieval}} and analysis of {{Martian}} haze profiles}.{\BBCQ}
\newblock
\APACjournalVolNumPages{Journal of Geophysical Research}{111}{E9}{}.
\newblock
\begin{APACrefDOI} \doi{10.1029/2005JE002662} \end{APACrefDOI}
\PrintBackRefs{\CurrentBib}

\bibitem [\protect \citeauthoryear {%
Murchie%
\ \protect \BOthers {.}}{%
Murchie%
\ \protect \BOthers {.}}{%
{\protect \APACyear {2007}}%
}]{%
murchie_CRISM_2007}
\APACinsertmetastar {%
murchie_CRISM_2007}%
\begin{APACrefauthors}%
Murchie, S.%
, Arvidson, R.%
, Bedini, P.%
, Beisser, K.%
, Bibring, J\BHBI P.%
, Bishop, J.%
\BDBL {}Wolff, M.%
\end{APACrefauthors}%
\unskip\
\newblock
\APACrefYearMonthDay{2007}{}{}.
\newblock
{\BBOQ}\APACrefatitle {Compact {Reconnaissance} {Imaging} {Spectrometer} for
  {Mars} ({CRISM}) on {Mars} {Reconnaissance} {Orbiter} ({MRO})} {Compact
  {Reconnaissance} {Imaging} {Spectrometer} for {Mars} ({CRISM}) on {Mars}
  {Reconnaissance} {Orbiter} ({MRO})}.{\BBCQ}
\newblock
\APACjournalVolNumPages{Journal of Geophysical Research: Planets}{112}{E5}{}.
\newblock
\begin{APACrefDOI} \doi{10.1029/2006JE002682} \end{APACrefDOI}
\PrintBackRefs{\CurrentBib}

\bibitem [\protect \citeauthoryear {%
Navarro%
, Forget%
, Millour%
\BCBL {}\ \BBA {} Greybush%
}{%
Navarro%
\ \protect \BOthers {.}}{%
{\protect \APACyear {2014}}%
}]{%
navarro_2014}
\APACinsertmetastar {%
navarro_2014}%
\begin{APACrefauthors}%
Navarro, T.%
, Forget, F.%
, Millour, E.%
\BCBL {}\ \BBA {} Greybush, S\BPBI J.%
\end{APACrefauthors}%
\unskip\
\newblock
\APACrefYearMonthDay{2014}{}{}.
\newblock
{\BBOQ}\APACrefatitle {Detection of Detached Dust Layers in the {{Martian}}
  Atmosphere from Their Thermal Signature Using Assimilation} {Detection of
  detached dust layers in the {{Martian}} atmosphere from their thermal
  signature using assimilation}.{\BBCQ}
\newblock
\APACjournalVolNumPages{Geophysical Research Letters}{41}{19}{6620-6626}.
\newblock
\begin{APACrefDOI} \doi{10.1002/2014GL061377} \end{APACrefDOI}
\PrintBackRefs{\CurrentBib}

\bibitem [\protect \citeauthoryear {%
Neary%
\ \protect \BOthers {.}}{%
Neary%
\ \protect \BOthers {.}}{%
{\protect \APACyear {2019}}%
}]{%
neary_2019}
\APACinsertmetastar {%
neary_2019}%
\begin{APACrefauthors}%
Neary, L.%
, Daerden, F.%
, Aoki, S.%
, Whiteway, J.%
, Clancy, R.%
, Smith, M.%
\BDBL {}Vandaele, A.%
\end{APACrefauthors}%
\unskip\
\newblock
\APACrefYearMonthDay{2019}{{\APACmonth{09}}}{}.
\newblock
{\BBOQ}\APACrefatitle {Explanation for the Increase in High Altitude Water on
  {{Mars}} Observed by {{NOMAD}} during the 2018 Global Dust Storm}
  {Explanation for the increase in high altitude water on {{Mars}} observed by
  {{NOMAD}} during the 2018 global dust storm}.{\BBCQ}
\newblock
\APACjournalVolNumPages{Geophysical Research Letters}{}{}{}.
\newblock
\begin{APACrefDOI} \doi{10.1029/2019GL084354} \end{APACrefDOI}
\PrintBackRefs{\CurrentBib}

\bibitem [\protect \citeauthoryear {%
Plane%
\ \protect \BOthers {.}}{%
Plane%
\ \protect \BOthers {.}}{%
{\protect \APACyear {2018}}%
}]{%
plane_2018}
\APACinsertmetastar {%
plane_2018}%
\begin{APACrefauthors}%
Plane, J\BPBI M\BPBI C.%
, Carrillo-Sanchez, J\BPBI D.%
, Mangan, T\BPBI P.%
, Crismani, M\BPBI M\BPBI J.%
, Schneider, N\BPBI M.%
\BCBL {}\ \BBA {} M{\"a}{\"a}tt{\"a}nen, A.%
\end{APACrefauthors}%
\unskip\
\newblock
\APACrefYearMonthDay{2018}{}{}.
\newblock
{\BBOQ}\APACrefatitle {Meteoric {{Metal Chemistry}} in the {{Martian
  Atmosphere}}} {Meteoric {{Metal Chemistry}} in the {{Martian
  Atmosphere}}}.{\BBCQ}
\newblock
\APACjournalVolNumPages{Journal of Geophysical Research:
  Planets}{123}{3}{695-707}.
\newblock
\begin{APACrefDOI} \doi{10.1002/2017JE005510} \end{APACrefDOI}
\PrintBackRefs{\CurrentBib}

\bibitem [\protect \citeauthoryear {%
Richardson%
}{%
Richardson%
}{%
{\protect \APACyear {2002}}%
}]{%
richardson_2002}
\APACinsertmetastar {%
richardson_2002}%
\begin{APACrefauthors}%
Richardson, M\BPBI I.%
\end{APACrefauthors}%
\unskip\
\newblock
\APACrefYearMonthDay{2002}{}{}.
\newblock
{\BBOQ}\APACrefatitle {Water ice clouds in the {Martian} atmosphere: {General}
  circulation model experiments with a simple cloud scheme} {Water ice clouds
  in the {Martian} atmosphere: {General} circulation model experiments with a
  simple cloud scheme}.{\BBCQ}
\newblock
\APACjournalVolNumPages{Journal of Geophysical Research}{107}{E9}{}.
\newblock
\begin{APACrefDOI} \doi{10.1029/2001JE001804} \end{APACrefDOI}
\PrintBackRefs{\CurrentBib}

\bibitem [\protect \citeauthoryear {%
S{\'a}nchez-Lavega%
, del R{\'i}o-Gaztelurrutia%
, Hern{\'a}ndez-Bernal%
\BCBL {}\ \BBA {} Delcroix%
}{%
S{\'a}nchez-Lavega%
\ \protect \BOthers {.}}{%
{\protect \APACyear {2019}}%
}]{%
sanchez-lavega_2019}
\APACinsertmetastar {%
sanchez-lavega_2019}%
\begin{APACrefauthors}%
S{\'a}nchez-Lavega, A.%
, del R{\'i}o-Gaztelurrutia, T.%
, Hern{\'a}ndez-Bernal, J.%
\BCBL {}\ \BBA {} Delcroix, M.%
\end{APACrefauthors}%
\unskip\
\newblock
\APACrefYearMonthDay{2019}{}{}.
\newblock
{\BBOQ}\APACrefatitle {The {{Onset}} and {{Growth}} of the 2018 {{Martian
  Global Dust Storm}}} {The {{Onset}} and {{Growth}} of the 2018 {{Martian
  Global Dust Storm}}}.{\BBCQ}
\newblock
\APACjournalVolNumPages{Geophysical Research Letters}{46}{11}{6101-6108}.
\newblock
\begin{APACrefDOI} \doi{10.1029/2019GL083207} \end{APACrefDOI}
\PrintBackRefs{\CurrentBib}

\bibitem [\protect \citeauthoryear {%
Smith%
}{%
Smith%
}{%
{\protect \APACyear {2004}}%
}]{%
smith_2004}
\APACinsertmetastar {%
smith_2004}%
\begin{APACrefauthors}%
Smith, M\BPBI D.%
\end{APACrefauthors}%
\unskip\
\newblock
\APACrefYearMonthDay{2004}{{\APACmonth{01}}}{}.
\newblock
{\BBOQ}\APACrefatitle {Interannual Variability in {{TES}} Atmospheric
  Observations of {{Mars}} during 1999\textendash{}2003} {Interannual
  variability in {{TES}} atmospheric observations of {{Mars}} during
  1999\textendash{}2003}.{\BBCQ}
\newblock
\APACjournalVolNumPages{Icarus}{167}{1}{148--165}.
\newblock
\begin{APACrefDOI} \doi{10.1016/j.icarus.2003.09.010} \end{APACrefDOI}
\PrintBackRefs{\CurrentBib}

\bibitem [\protect \citeauthoryear {%
Smith%
}{%
Smith%
}{%
{\protect \APACyear {2019}}%
}]{%
smith_2019}
\APACinsertmetastar {%
smith_2019}%
\begin{APACrefauthors}%
Smith, M\BPBI D.%
\end{APACrefauthors}%
\unskip\
\newblock
\APACrefYearMonthDay{2019}{{\APACmonth{11}}}{}.
\newblock
{\BBOQ}\APACrefatitle {{{THEMIS}} Observations of the 2018 {{Mars}} Global Dust
  Storm} {{{THEMIS}} observations of the 2018 {{Mars}} global dust
  storm}.{\BBCQ}
\newblock
\APACjournalVolNumPages{Journal of Geophysical Research: Planets}{}{}{}.
\newblock
\begin{APACrefDOI} \doi{10.1029/2019JE006107} \end{APACrefDOI}
\PrintBackRefs{\CurrentBib}

\bibitem [\protect \citeauthoryear {%
Smith%
, Wolff%
, Clancy%
, Kleinb\"ohl%
\BCBL {}\ \BBA {} Murchie%
}{%
Smith%
\ \protect \BOthers {.}}{%
{\protect \APACyear {2013}}%
}]{%
smith_2013}
\APACinsertmetastar {%
smith_2013}%
\begin{APACrefauthors}%
Smith, M\BPBI D.%
, Wolff, M\BPBI J.%
, Clancy, R\BPBI T.%
, Kleinb\"ohl, A.%
\BCBL {}\ \BBA {} Murchie, S\BPBI L.%
\end{APACrefauthors}%
\unskip\
\newblock
\APACrefYearMonthDay{2013}{{\APACmonth{02}}}{}.
\newblock
{\BBOQ}\APACrefatitle {Vertical distribution of dust and water ice aerosols
  from {CRISM} limb-geometry observations} {Vertical distribution of dust and
  water ice aerosols from {CRISM} limb-geometry observations}.{\BBCQ}
\newblock
\APACjournalVolNumPages{Journal of Geophysical Research:
  Planets}{118}{2}{321--334}.
\newblock
\begin{APACrefDOI} \doi{10.1002/jgre.20047} \end{APACrefDOI}
\PrintBackRefs{\CurrentBib}

\bibitem [\protect \citeauthoryear {%
Szantai%
\ \protect \BOthers {.}}{%
Szantai%
\ \protect \BOthers {.}}{%
{\protect \APACyear {2019}}%
}]{%
szantai_2019}
\APACinsertmetastar {%
szantai_2019}%
\begin{APACrefauthors}%
Szantai, A.%
, Audouard, J.%
, Forget, F.%
, Olsen, K\BPBI S.%
, Gondet, B.%
, Millour, E.%
\BDBL {}Bibring, J\BHBI P.%
\end{APACrefauthors}%
\unskip\
\newblock
\APACrefYearMonthDay{2019}{{\APACmonth{04}}}{}.
\newblock
{\BBOQ}\APACrefatitle {Martian Cloud Climatology and Life Cycle Extracted from
  {{Mars Express OMEGA}} Spectral Images} {Martian cloud climatology and life
  cycle extracted from {{Mars Express OMEGA}} spectral images}.{\BBCQ}
\newblock
\APACjournalVolNumPages{arXiv e-prints}{}{}{arXiv:1904.06422}.
\PrintBackRefs{\CurrentBib}

\bibitem [\protect \citeauthoryear {%
Toon%
\ \BBA {} Ackerman%
}{%
Toon%
\ \BBA {} Ackerman%
}{%
{\protect \APACyear {1981}}%
}]{%
toon_1981}
\APACinsertmetastar {%
toon_1981}%
\begin{APACrefauthors}%
Toon, O\BPBI B.%
\BCBT {}\ \BBA {} Ackerman, T\BPBI P.%
\end{APACrefauthors}%
\unskip\
\newblock
\APACrefYearMonthDay{1981}{{\APACmonth{10}}}{}.
\newblock
{\BBOQ}\APACrefatitle {Algorithms for the calculation of scattering by
  stratified spheres} {Algorithms for the calculation of scattering by
  stratified spheres}.{\BBCQ}
\newblock
\APACjournalVolNumPages{Applied Optics}{20}{20}{3657}.
\newblock
\begin{APACrefDOI} \doi{10.1364/AO.20.003657} \end{APACrefDOI}
\PrintBackRefs{\CurrentBib}

\bibitem [\protect \citeauthoryear {%
Trokhimovskiy%
\ \protect \BOthers {.}}{%
Trokhimovskiy%
\ \protect \BOthers {.}}{%
{\protect \APACyear {2015}}%
}]{%
trokhimovskiy_2015}
\APACinsertmetastar {%
trokhimovskiy_2015}%
\begin{APACrefauthors}%
Trokhimovskiy, A.%
, Korablev, O.%
, Ivanov, Y\BPBI S.%
, Siniyavsky, I\BPBI I.%
, Fedorova, A.%
, Stepanov, A\BPBI V.%
\BDBL {}Montmessin, F.%
\end{APACrefauthors}%
\unskip\
\newblock
\APACrefYearMonthDay{2015}{{\APACmonth{09}}}{}.
\newblock
{\BBOQ}\APACrefatitle {Middle-infrared echelle cross-dispersion spectrometer
  {ACS}-{MIR} for the {ExoMars} {Trace} {Gas} {Orbiter}} {Middle-infrared
  echelle cross-dispersion spectrometer {ACS}-{MIR} for the {ExoMars} {Trace}
  {Gas} {Orbiter}}.{\BBCQ}
\newblock
\BIn{} \APACrefbtitle {Infrared Remote Sensing and Instrumentation {XXIII}}
  {Infrared remote sensing and instrumentation {XXIII}}\ (\BVOL\ 9608,
  \BPG~960808).
\newblock
\begin{APACrefDOI} \doi{10.1117/12.2190359} \end{APACrefDOI}
\PrintBackRefs{\CurrentBib}

\bibitem [\protect \citeauthoryear {%
Vals%
, Forget%
, Spiga%
\BCBL {}\ \BBA {} Millour%
}{%
Vals%
\ \protect \BOthers {.}}{%
{\protect \APACyear {2018}}%
}]{%
vals_2018}
\APACinsertmetastar {%
vals_2018}%
\begin{APACrefauthors}%
Vals, M.%
, Forget, F.%
, Spiga, A.%
\BCBL {}\ \BBA {} Millour, E.%
\end{APACrefauthors}%
\unskip\
\newblock
\APACrefYearMonthDay{2018}{{\APACmonth{09}}}{}.
\newblock
{\BBOQ}\APACrefatitle {Impact of the Refinement of the Vertical Resolution on
  the Simulation of the Water Cycle by the Martian {{LMD Global Climate
  Model}}} {Impact of the refinement of the vertical resolution on the
  simulation of the water cycle by the martian {{LMD Global Climate
  Model}}}.{\BBCQ}
\newblock
\BIn{} (\BVOL~12, \BPG~EPSC2018-847).
\PrintBackRefs{\CurrentBib}

\bibitem [\protect \citeauthoryear {%
Vandaele%
\ \protect \BOthers {.}}{%
Vandaele%
\ \protect \BOthers {.}}{%
{\protect \APACyear {2019}}%
}]{%
nature_hdo}
\APACinsertmetastar {%
nature_hdo}%
\begin{APACrefauthors}%
Vandaele, A\BPBI C.%
, Korablev, O.%
, Daerden, F.%
, Aoki, S.%
, Thomas, I\BPBI R.%
, Altieri, F.%
\BDBL {}{ACS Science Team}%
\end{APACrefauthors}%
\unskip\
\newblock
\APACrefYearMonthDay{2019}{{\APACmonth{04}}}{}.
\newblock
{\BBOQ}\APACrefatitle {Martian dust storm impact on atmospheric {H}$_2${O} and
  {D}/{H} observed by {ExoMars} {Trace} {Gas} {Orbiter}} {Martian dust storm
  impact on atmospheric {H}$_2${O} and {D}/{H} observed by {ExoMars} {Trace}
  {Gas} {Orbiter}}.{\BBCQ}
\newblock
\APACjournalVolNumPages{Nature}{568}{}{521--525}.
\newblock
\begin{APACrefDOI} \doi{10.1038/s41586-019-1097-3} \end{APACrefDOI}
\PrintBackRefs{\CurrentBib}

\bibitem [\protect \citeauthoryear {%
Vandaele%
\ \protect \BOthers {.}}{%
Vandaele%
\ \protect \BOthers {.}}{%
{\protect \APACyear {2018}}%
}]{%
vandaele_nomad_2018}
\APACinsertmetastar {%
vandaele_nomad_2018}%
\begin{APACrefauthors}%
Vandaele, A\BPBI C.%
, Lopez-Moreno, J\BHBI J.%
, Patel, M\BPBI R.%
, Bellucci, G.%
, Daerden, F.%
, Ristic, B.%
\BDBL {}{the NOMAD Team}%
\end{APACrefauthors}%
\unskip\
\newblock
\APACrefYearMonthDay{2018}{{\APACmonth{08}}}{}.
\newblock
{\BBOQ}\APACrefatitle {{NOMAD}, an {Integrated} {Suite} of {Three}
  {Spectrometers} for the {ExoMars} {Trace} {Gas} {Mission}: {Technical}
  {Description}, {Science} {Objectives} and {Expected} {Performance}} {{NOMAD},
  an {Integrated} {Suite} of {Three} {Spectrometers} for the {ExoMars} {Trace}
  {Gas} {Mission}: {Technical} {Description}, {Science} {Objectives} and
  {Expected} {Performance}}.{\BBCQ}
\newblock
\APACjournalVolNumPages{Space Science Reviews}{214}{5}{}.
\newblock
\begin{APACrefDOI} \doi{10.1007/s11214-018-0517-2} \end{APACrefDOI}
\PrintBackRefs{\CurrentBib}

\bibitem [\protect \citeauthoryear {%
Vincendon%
, Pilorget%
, Gondet%
, Murchie%
\BCBL {}\ \BBA {} Bibring%
}{%
Vincendon%
\ \protect \BOthers {.}}{%
{\protect \APACyear {2011}}%
}]{%
vincendon_2011}
\APACinsertmetastar {%
vincendon_2011}%
\begin{APACrefauthors}%
Vincendon, M.%
, Pilorget, C.%
, Gondet, B.%
, Murchie, S.%
\BCBL {}\ \BBA {} Bibring, J\BHBI P.%
\end{APACrefauthors}%
\unskip\
\newblock
\APACrefYearMonthDay{2011}{{\APACmonth{11}}}{}.
\newblock
{\BBOQ}\APACrefatitle {New near-{IR} observations of mesospheric {CO}$_2$ and
  {H}$_2${O} clouds on {Mars}} {New near-{IR} observations of mesospheric
  {CO}$_2$ and {H}$_2${O} clouds on {Mars}}.{\BBCQ}
\newblock
\APACjournalVolNumPages{Journal of Geophysical Research}{116}{}{E00J02}.
\newblock
\begin{APACrefDOI} \doi{10.1029/2011JE003827} \end{APACrefDOI}
\PrintBackRefs{\CurrentBib}

\bibitem [\protect \citeauthoryear {%
Virtanen%
\ \protect \BOthers {.}}{%
Virtanen%
\ \protect \BOthers {.}}{%
{\protect \APACyear {2019}}%
}]{%
scipy_preprint}
\APACinsertmetastar {%
scipy_preprint}%
\begin{APACrefauthors}%
Virtanen, P.%
, Gommers, R.%
, Oliphant, T\BPBI E.%
, Haberland, M.%
, Reddy, T.%
, Cournapeau, D.%
\BDBL {}Contributors, S\BPBI \BPBI .%
\end{APACrefauthors}%
\unskip\
\newblock
\APACrefYearMonthDay{2019}{{\APACmonth{07}}}{}.
\newblock
{\BBOQ}\APACrefatitle {{{SciPy}} 1.0--{{Fundamental Algorithms}} for
  {{Scientific Computing}} in {{Python}}} {{{SciPy}} 1.0--{{Fundamental
  Algorithms}} for {{Scientific Computing}} in {{Python}}}.{\BBCQ}
\newblock
\APACjournalVolNumPages{arXiv:1907.10121 [physics]}{}{}{}.
\PrintBackRefs{\CurrentBib}

\bibitem [\protect \citeauthoryear {%
Wang%
\ \BBA {} Richardson%
}{%
Wang%
\ \BBA {} Richardson%
}{%
{\protect \APACyear {2015}}%
}]{%
wang_richardson_2015}
\APACinsertmetastar {%
wang_richardson_2015}%
\begin{APACrefauthors}%
Wang, H.%
\BCBT {}\ \BBA {} Richardson, M\BPBI I.%
\end{APACrefauthors}%
\unskip\
\newblock
\APACrefYearMonthDay{2015}{{\APACmonth{05}}}{}.
\newblock
{\BBOQ}\APACrefatitle {The origin, evolution, and trajectory of large dust
  storms on {Mars} during {Mars} years 24–30 (1999–2011)} {The origin,
  evolution, and trajectory of large dust storms on {Mars} during {Mars} years
  24–30 (1999–2011)}.{\BBCQ}
\newblock
\APACjournalVolNumPages{Icarus}{251}{}{112--127}.
\newblock
\begin{APACrefDOI} \doi{10.1016/j.icarus.2013.10.033} \end{APACrefDOI}
\PrintBackRefs{\CurrentBib}

\bibitem [\protect \citeauthoryear {%
Wilson%
, Lewis%
, Montabone%
\BCBL {}\ \BBA {} Smith%
}{%
Wilson%
\ \protect \BOthers {.}}{%
{\protect \APACyear {2008}}%
}]{%
wilson_2008}
\APACinsertmetastar {%
wilson_2008}%
\begin{APACrefauthors}%
Wilson, R\BPBI J.%
, Lewis, S\BPBI R.%
, Montabone, L.%
\BCBL {}\ \BBA {} Smith, M\BPBI D.%
\end{APACrefauthors}%
\unskip\
\newblock
\APACrefYearMonthDay{2008}{}{}.
\newblock
{\BBOQ}\APACrefatitle {Influence of water ice clouds on {Martian} tropical
  atmospheric temperatures} {Influence of water ice clouds on {Martian}
  tropical atmospheric temperatures}.{\BBCQ}
\newblock
\APACjournalVolNumPages{Geophysical Research Letters}{35}{7}{}.
\newblock
\begin{APACrefDOI} \doi{10.1029/2007GL032405} \end{APACrefDOI}
\PrintBackRefs{\CurrentBib}

\bibitem [\protect \citeauthoryear {%
Wolff%
\ \BBA {} Clancy%
}{%
Wolff%
\ \BBA {} Clancy%
}{%
{\protect \APACyear {2003}}%
}]{%
wolff_2003}
\APACinsertmetastar {%
wolff_2003}%
\begin{APACrefauthors}%
Wolff, M\BPBI J.%
\BCBT {}\ \BBA {} Clancy, R\BPBI T.%
\end{APACrefauthors}%
\unskip\
\newblock
\APACrefYearMonthDay{2003}{}{}.
\newblock
{\BBOQ}\APACrefatitle {Constraints on the Size of {{Martian}} Aerosols from
  {{Thermal Emission Spectrometer}} Observations} {Constraints on the size of
  {{Martian}} aerosols from {{Thermal Emission Spectrometer}}
  observations}.{\BBCQ}
\newblock
\APACjournalVolNumPages{Journal of Geophysical Research: Planets}{108}{E9}{}.
\newblock
\begin{APACrefDOI} \doi{10.1029/2003JE002057} \end{APACrefDOI}
\PrintBackRefs{\CurrentBib}

\bibitem [\protect \citeauthoryear {%
Wolff%
\ \protect \BOthers {.}}{%
Wolff%
\ \protect \BOthers {.}}{%
{\protect \APACyear {2019}}%
}]{%
wolff_2019}
\APACinsertmetastar {%
wolff_2019}%
\begin{APACrefauthors}%
Wolff, M\BPBI J.%
, Clancy, R\BPBI T.%
, Kahre, M\BPBI A.%
, Haberle, R\BPBI M.%
, Forget, F.%
, Cantor, B\BPBI A.%
\BCBL {}\ \BBA {} Malin, M\BPBI C.%
\end{APACrefauthors}%
\unskip\
\newblock
\APACrefYearMonthDay{2019}{{\APACmonth{11}}}{}.
\newblock
{\BBOQ}\APACrefatitle {Mapping Water Ice Clouds on {{Mars}} with
  {{MRO}}/{{MARCI}}} {Mapping water ice clouds on {{Mars}} with
  {{MRO}}/{{MARCI}}}.{\BBCQ}
\newblock
\APACjournalVolNumPages{Icarus}{332}{}{24--49}.
\newblock
\begin{APACrefDOI} \doi{10.1016/j.icarus.2019.05.041} \end{APACrefDOI}
\PrintBackRefs{\CurrentBib}

\bibitem [\protect \citeauthoryear {%
{Wolff}%
\ \protect \BOthers {.}}{%
{Wolff}%
\ \protect \BOthers {.}}{%
{\protect \APACyear {2017}}%
}]{%
wolff_2017}
\APACinsertmetastar {%
wolff_2017}%
\begin{APACrefauthors}%
{Wolff}, M\BPBI J.%
, {Lop{\'e}z-Valverde}, M.%
, {Madeleine}, J\BHBI B.%
, {Wilson}, R\BPBI J.%
, {Smith}, M\BPBI D.%
, {Fouchet}, T.%
\BCBL {}\ \BBA {} {Delory}, G\BPBI T.%
\end{APACrefauthors}%
\unskip\
\newblock
\APACrefYearMonthDay{2017}{}{}.
\newblock
{\BBOQ}\APACrefatitle {{Radiative Process: Techniques and Applications}}
  {{Radiative Process: Techniques and Applications}}.{\BBCQ}
\newblock
\BIn{} R\BPBI M.~Haberle\ \BOthers {.}\ (\BEDS), \APACrefbtitle {The atmosphere
  and climate of {Mars}} {The atmosphere and climate of {Mars}}\ (\BPG~76-105).
\newblock
\APACaddressPublisher{}{{Cambridge} {University} {Press}}.
\newblock
\begin{APACrefDOI} \doi{10.1017/9781139060172.006} \end{APACrefDOI}
\PrintBackRefs{\CurrentBib}

\bibitem [\protect \citeauthoryear {%
Zurek%
\ \BBA {} Martin%
}{%
Zurek%
\ \BBA {} Martin%
}{%
{\protect \APACyear {1993}}%
}]{%
zurek_1993}
\APACinsertmetastar {%
zurek_1993}%
\begin{APACrefauthors}%
Zurek, R\BPBI W.%
\BCBT {}\ \BBA {} Martin, L\BPBI J.%
\end{APACrefauthors}%
\unskip\
\newblock
\APACrefYearMonthDay{1993}{{\APACmonth{02}}}{}.
\newblock
{\BBOQ}\APACrefatitle {Interannual variability of planet-encircling dust storms
  on {Mars}} {Interannual variability of planet-encircling dust storms on
  {Mars}}.{\BBCQ}
\newblock
\APACjournalVolNumPages{Journal of Geophysical Research:
  Planets}{98}{E2}{3247--3259}.
\newblock
\begin{APACrefDOI} \doi{10.1029/92JE02936} \end{APACrefDOI}
\PrintBackRefs{\CurrentBib}

\end{thebibliography}

\end{document}